\documentclass{ar-1col-S2O}
\usepackage{sidecap}

\usepackage[comma]{natbib}
\usepackage{url}
\begin{document}

% Page header
\markboth{Kar and DiCarlo}{Mechanisms of Primate Object Recognition}

% Title

% Previous title
%\title{Integrated Set of Neural Mechanisms Underlying Object Recognition in Primates}

% Alternative new title:

%\title{The quest for a sensory computable, mechanistic, anatomically referenced, and testable (SMART) understanding of primate object recognition}
\title{The Quest for an Integrated Set of Neural Mechanisms Underlying Object Recognition in Primates}

%Authors, affiliations address.
\author{Kohitij Kar$^{1}$, and James J. DiCarlo$^{2}$
\affil{$^1$ York University, Department of Biology, Centre for Vision Research, Toronto, Canada; email: k0h1t1j@yorku.ca}
\affil{$^2$ McGovern Institute for Brain Research, Dept. of Brain and Cognitive Sciences, Massachusetts Institute of Technology, Cambridge, USA; email: dicarlo@mit.edu}
}

%Abstract
\begin{abstract}
Visual object recognition — the behavioral ability to rapidly and accurately categorize many visually encountered objects -- is core to primate cognition. This behavioral capability is algorithmically impressive because of the myriad identity-preserving viewpoints and scenes that dramatically change the visual image produced by the same object. Until recently, the brain mechanisms that support that capability were deeply mysterious. However, over the last decade, this scientific mystery has been illuminated by the discovery and development of brain-inspired, image-computable, artificial neural network (ANN) systems that rival primates in this behavioral feat. Apart from fundamentally changing the landscape of artificial intelligence (AI), modified versions of these ANN systems are the current leading scientific hypotheses of an integrated set of mechanisms in the primate ventral visual stream that support object recognition. What separates brain-mapped versions of these systems from prior conceptual models is that they are Sensory-computable, Mechanistic, Anatomically Referenced, and Testable (SMART). Here, we review and provide perspective on the brain mechanisms that the currently leading SMART models address. We review the empirical brain and behavioral alignment successes and failures of those current models. Given ongoing advances in neurobehavioral measurements and AI, we discuss the next frontiers for even more accurate mechanistic understanding. And we outline the likely applications of that SMART-model-based understanding.

\end{abstract}

%Keywords, etc.

\maketitle

%Table of Contents
\tableofcontents

% Heading 1
\section{Introduction}

Primates can rapidly recognize and report multiple details about real-world objects in their field of view, despite the potentially infinite variation that an image of an object might present to the eyes \citep{rajalingham2015comparison, rajalingham2018large}. A decade ago, neuroscientists had already successfully probed the primate brain's visual processing pathways to infer a conceptual understanding of object recognition (for a review, see \citealp[]{dicarlo2012does}). In particular, prior work had demonstrated the central role of the ventral visual cortical stream for processing the visual input at the center of gaze to support object recognition behaviors \citep{ungerleider1982two}.  And neural recordings along the ventral visual pathway \citep{logothetis1995shape, hung2005fast, majaj2015simple} had demonstrated the presence of critical neural circuits responsible for enabling primates with their remarkable object recognition capabilities. Furthermore, anatomically constrained, specialized circuits had been discovered that exhibit selectively for specific visual objects, and image statistics \citep{gross1972visual, tanaka1996inferotemporal, op2001inferotemporal}. For instance, a group of neurons in the inferior temporal (IT) cortex that is more responsive to faces compared to other objects \citep{kanwisher1997fusiform,tsao2006cortical,leopold2006norm} has been casually linked to face perception \citep{sadagopan2017causal,schalk2017facephenes}. Similarly, various other lines of research have also discovered other functional topographies in the visual cortex \citep{popivanov2014heterogeneous,lafer2013parallel}. However, how an end-to-end model that could receive an image as an input and accurately transform this pixel-level information to perform visual object recognition tasks while also reproducing  the ventral stream's intermediate  solutions remained unsolved \citep{dicarlo2012does}.

Over decades, a relatively small cadre of the computer vision community \citep{lecun1989backpropagation, lecun1995convolutional, rumelhart1986learning} and the computational neuroscience vision community \citep{riesenhuber1999hierarchical, pinto2009high} worked to stay close to the anatomy of the ventral stream.  Beginning in 2012, vision system builders in this ventral stream inspired lineage -- when fueled by more powerful computers and larger data sets \citep{russakovsky2015imagenet} -- began to make remarkable strides in developing machine vision systems \citep{krizhevsky2012imagenet,he2016deep} that could solve this very hard problem of visual object recognition with near human-level accuracy. At nearly the same time, the visual neuroscience community began to show that systems built in this manner were by far the empirically leading \emph{scientific} models of the primate brain mechanisms that support object recognition \citep{yamins2013hierarchical,yamins2014performance, khaligh2014deep, cadieu2014deep}. 

\textbf{This review is organized around reproducible models of the integrated set of mechanisms supporting object recognition}, their evaluation, successes and shortcomings, and how existing and future neural and behavioral data can help guide the development of the next generation of such models. 
More specifically, our review is focused on the understanding that the field has achieved thus far in the non-human primate system.  We explain why we think (see Box 1) that this understanding will — with some modifications — generalize to mechanisms at work in the human brain. 

Before reviewing this progress, we set some basic premises: what do we mean by object recognition (1.1)?, what do we mean by an \emph{understanding} of object recognition (1.2)?, and more specifically -- what is a \emph{mechanistic} understanding of object recognition (1.3)? The purpose of this clarification is not to discourage or critique other approaches and viewpoints but rather to define the scope and perspective of this article.  In addition, the reader should note that object recognition is only a starting point in studying visual intelligence.  As such, we synthesize our discussions toward the goal of discovering the algorithms of visual intelligence more generally. 

\begin{textbox}[h]
  \section{Box 1: Rationale for the non-human primate animal model}  

In this article, we provide a macaque-centric review of the brain mechanisms of object recognition that we believe readily apply to humans. We reason that we need an animal model that meets three criteria to successfully probe the neural mechanisms of visual object recognition in humans. First, to interchangeably test the animal model and human subjects, the animals should have human-level perceptual capabilities assessed by quantitative metrics\citep{rajalingham2018large, rajalingham2015comparison}. Second, it should be possible to conduct fine-grained neural measurements \citep{kar2019evidence,trautmann2023large} and targeted causal perturbations to interrogate brain circuits \citep{kar2021fast, azadi2023image, rajalingham2021chronically} that are not feasible in humans. Third, inferences made on the animals’ neural mechanisms should be relevant for humans due to evolutionary proximity \citep{perelman2011molecular} and established homologies in brain areas and behavior (behavioral similarities shown in \textbf{Figure 3}). The rhesus macaque, with its ventral visual cortex that supports human-like object processing, meets all of these criteria.  Thus, an accurate model of the neural mechanisms of monkey behavior will likely readily generalize to an understanding of the homologous human brain systems. In the rest of the article, when we write “primate”, we mean human and non-human primates. For more human-centric reviews, we refer the reader to \citealp{peters2021capturing}.

\end{textbox}

\subsection{What do we mean by object recognition?}

When a human observer encounters a visual scene (e.g., walking into a room), they quickly infer many things about the world content of that scene. To the extent that each report agrees with the underlying physical content of the world, we say that those inferences are accurate. The variables of the true physical content of the world such as the number of objects, the shape of each object, the category of each object, the pose of each object relative to the viewer, etc., are called the ``latent" content because the discrete or scalar values of those variables are not explicitly available to (i.e., they are hidden from the) the perceptual system.  For visual systems, such values must be inferred only from spatiotemporal patterns of photons striking the eyes. Remarkably, however, human reports of these values are highly accurate even with just single views of the scene – aka single “images.” For instance, if you look briefly at the example image shown in \textbf{Figure 1}, you are likely able to answer many questions about the image — did you see a bird or a cat? did you see an owl or an osprey? was the bird left or right of the fixation cross? was the bird behind or in front of a branch? was that a novel or a familiar bird? Is the image pleasant or threatening?  etc. Typically, the study of object recognition within a scene is focused on primates' ability to determine the specific identity and category of a dominant foreground object (related to the first two questions \textbf{Figure 1A}), but the broader domain of visual intelligence includes all of those questions and many more. 

\begin{figure}[h]
\includegraphics[width=10.66 cm]{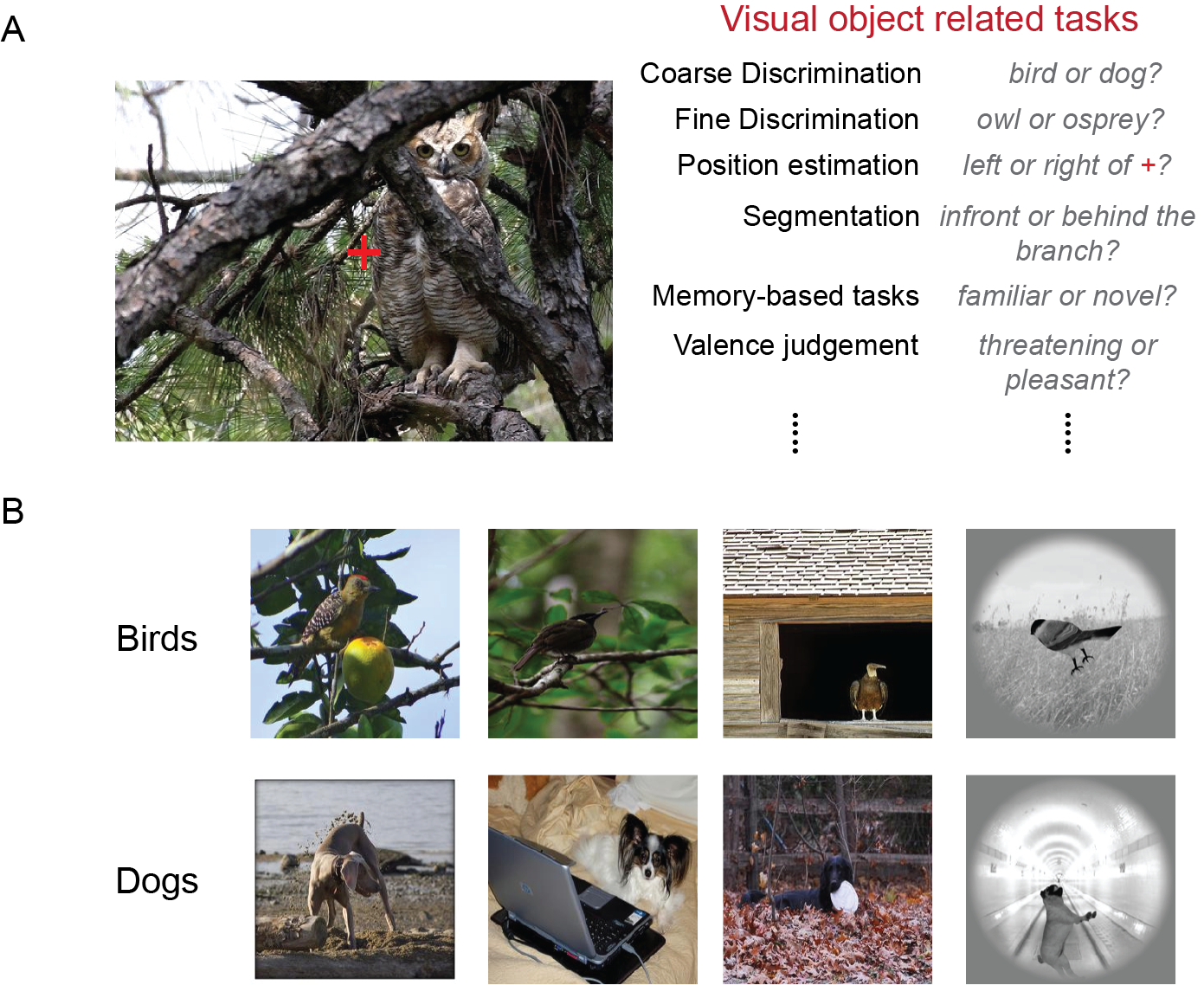}
\caption{\textbf{Probing Visual Object Perception Through Diverse Behavioral Tasks.}
\textbf{A.} The left panel shows an image -- a two-dimensional pixel grid of RGB luminances -- comparable to a photoreceptor-transduced spatial pattern of physical (photon) energy striking your retinae just after you turn your head to look up. On the right, a series of visual object-related perceptual report tasks are listed, highlighting the multifaceted ways to investigate a subject's perceptual inferences about the latent content of the world from just the image alone. These tasks span from coarse-category distinctions (such as discerning between the presence of a bird or a cat in the image) to more complex evaluations linked to memory and emotional responses (for instance, determining the familiarity of the bird or assessing its emotional valence). \textbf{B.} Object recognition is algorithmically challenging because the same object category (i.e., the same type of latent cause) can generate a potentially infinite set of images and successful behavior depends on inferring the presence of that object for any such image. Two rows of examples of images from the category bird (top row) and dog (bottom row) are shown to demonstrate this challenge conceptually.}
\label{fig1}
\end{figure}

To ask if we are making scientific progress on understanding the mechanisms (see 1.2) that enable us to perform object recognition, it is essential to operationalize a starting set of tasks -- both to assess and characterize biological performance and the performance of computational models that aim to explain how that biology works. DiCarlo et al. \citeyear{dicarlo2012does}, proposed \textbf{“core object recognition”} as a starting point in that effort. By definition, core object recognition confines the visual intelligence challenge to the processing of images presented within the subject's central field of view (central 10 degrees of visual angle) and for a limited time ($<$ 200 ms). This operational definition was chosen because: it is known that human shape discrimination abilities are best at the center of gaze, that the ventral visual stream processing is dominated by the central 10 degrees \citep{ungerleider1982two,op2000spatial}, 200 msec  corresponds to the duration of fixation during natural viewing behavior \citep{nuthmann2017fixation, dicarlo2000form}, and object categorization performance at the center of gaze was known to be already remarkably accurate at this duration and even much shorter \citep{thorpe1996speed,potter1976short,keysers2001speed}. 

Having rationally operationalized the sensory input domain above (10 deg, 200 msec) there are still many ways one might operationally assay the perceptual contents of human or animal minds around objects.   For instance, one could “pre-cue” the subject about the types of objects to expect, and this could be done either explicitly (e.g., ``You will see either a bird or a dog next.") or implicitly (e.g., testing a block of many 'bird vs. dog' trials).  Indeed, effects of pre-cueing have been extensively pursued in the “attention” literature \citep{zhang2011object}. Human observers can also be asked to only report the object as a post-trial questionnaire or discrimination task. This article focuses on the post-cueing framework in which subjects enter each trial with many possible objects to entertain (typically at least 8), and the question of which object was present is asked immediately after a test image.  We consider this paradigm to put subjects in a default attentional mode in which ``spatial attention" \citep{maunsell2015neuronal} is at the center of the scene (it is implicitly pre-cued) and ``feature attention” \citep{maunsell2006feature} is also in a default mode in that the visual system can emphasize no single set of features due to the large number of potential objects and the associated complexity of features that must be handled to succeed in the task.  We do not mean to imply that spatial and feature attention phenomena should not be part of a complete understanding of visual processing and visual intelligence, but only that the mechanisms underlying those attentional phenomena are in reasonably natural default modes for most of the empirical studies we discuss below.   

As motivated above, core object recognition focuses specifically on the 100-200 ms viewing duration time scale, and hence, we review mechanisms that are most relevant for that time scale.  Longer viewing of images and videos will likely require additional mechanistic components beyond core recognition, including mechanisms related to directing eye movements. Beyond object category and identity, other object-related latent variables such as object size, position, rotation, color, and material properties not only affect human estimates of object identity (for more discussions see, \citealp{bracci2023understanding}), but are themselves variables of objects that humans must also often accurately infer and are within the scope of core object recognition. In addition, the values of other object-related latent variables such as object motion trajectory, velocity, etc., could impact “object-identity” estimates and are also within the scope of core recognition. 

In sum, the conceptual "output" of core object recognition is the contents of the subject's perceptual state causally induced by each image (e.g., the values of the set of object-related latent variables, above). Key operational measures of this output include the subject's behavioral reports of those contents, given a task paradigm (i.e., a way to trigger such reports). That is, we say that each image \emph{causes} a particular perceptual state and its associated behavioral reports in that presentation of an image of a cat will reliably produce the behavioral report of “cat” and removal of that image (e.g., presenting a full field gray image) will reliably eliminate that behavioral report.  

Our review is primarily focused on progress in understanding primate brain mechanisms that underlie core object recognition. Given the above definitions and paradigms, it should be clear that core object recognition is not the entirety of what one might want to call “object recognition,” and, it is certainly not all of visual perception and visual intelligence. Nevertheless, the progress outlined below suggests that -- somewhat fortuitously -- a very large fraction of human ability to estimate the values of object latent variables (above) and the underlying visual processing that underlies many tasks beyond object recognition can be understood via the computable models that come out of this “solve core object recognition first” approach.  

\subsection{What do we mean by an \emph{understanding} of core object recognition?}

Much of scientific understanding is in the form of reproducible models \citep{popper1934logic, Kuhn1962}, ideally coupled to robust theoretical frameworks. Thus, any understanding of core object recognition should minimally include models that can potentially explain and predict empirical patterns of behavior for \emph{any} image in the core recognition input domain (central 10 deg,  $<$200 msec).  The field does not agree on all model desiderata (e.g., compactness, explainability to others, etc.). Thus, the field does not fully agree on what comprises an “understanding."  In this review, we focus on models with four primary desiderata: 1) high reproducibility (i.e., models that, for any image, produce the same predictions in the hands of other scientists), 2) high empirical accuracy at the behavioral level (i.e., models whose predictions on new images tend to match the empirical observations of behavior; e.g., match the pattern of successes and failures over images, where success is defined with respect to ground truth objects that generated the test images), 3) brain-mapped mechanisms (at a particular level of resolution, defined below), and 4) high empirical accuracy at the neural level (i.e., the model predictions tend to match the empirical observations at the mapped level of resolution).

We note that a model does not need to meet all four desiderata to be useful. For example, models that meet desiderata 1 $\&$ 2 would contribute to cognitive science.  And models that meet desiderata 1, 3 $\&$ 4 would contribute to neuroscience. However, models of the integrated set of neural mechanisms that underlie core object recognition must ultimately meet all four desiderata. To meet desiderata (1), we focus on ``computable” models that define a precise procedure (usually specified in software) that can be readily shared with other scientists to produce the same model predictions in different laboratories. As such, computable models as defined here have very high reproducibility.  

In core visual object recognition (a behavioral capability), computable models must minimally take images (i.e., spatial patterns of photons) as input and produce behavioral reports in response to each image as output.  Models that can make predictions (e.g., behavioral report predictions) for \emph{any} given image are referred to as image-computable or, equivalently, sensory-computable models (See Box 2).  
%In our opinion, 
Image-computable models are scientifically crucial because they engage the full complexity of natural images, and they are precisely reproducible in the hands of other scientists and are thus independently testable \citep{yamins2016using}.  
%Not all vision researchers agree with this [ref=Bowers]

In sum, any sensory-computable model that accurately predicts the patterns of core object recognition behavior would, to us, constitute a potential causal scientific understanding of core object recognition.  We note that some view this as necessary, but not sufficient, for understanding. We are not opposed to that view, but we strongly oppose the view that such models are not even necessary (see \cite{schrimpf2018brain} for a discussion of this issue).  

We do not mean to imply that only one such model exists (actually, an infinite number exists).  Nor do we mean to imply that other model desiderata are not potentially useful. Indeed, we are particularly interested in models that are not only capable of explaining the behavioral pattern resulting from the sensory input but are also capable of explaining how different parts of the brain work together (aka the underlying neural mechanism) to produce those behavioral patterns at various levels of detail. Next, we elaborate on what we mean by "mechanistic" models of object recognition.  

%\subsection{What constitutes a mechanistic model of core object recognition?}

\subsection{What is a \textit{mechanistic} understanding?}

\begin{marginnote}[]
\entry{Image-computable model}{A machine executable system that can take any image as input and produce neural and/or behavioral predictions as outputs.}

\entry{Sensory-computable model} {The same concept (above), generalized to any sensory system.}
%(auditory/language, somatosensory, olfactory, etc.) }

\entry{ANN}{Artificial Neural Network: A machine executable system made up only of connected sets of weighted summation nodes ("neurons")}

%In this review we use the term ANN to refer to such systems generated by computer vision and AI researchers even if those system builders have no further necessity or intent to engage with neuroscience}

\entry{Deep ANN} {A multi-layer ANN, where the output of each layer of neurons provides most of the input to the next deeper layer.}

\entry{SMART models}{\textbf{S}ensory-computable, \textbf{M}echanistic, \textbf{A}natomically-\textbf{R}eferenced, \textbf{T}estable models.   May be built by neuroscientists or inherited from AI system builders and then modified and mapped to the brain.  
%Regardless, these models serve as mechanistic hypotheses of brain sub-systems and their supported behavioral capabilities.  
The current leading SMART models of primate core object recognition are image-computable deep convolutional ANNs derived from computer vision.  Each layer models a brain region or area (see Box 2).}

\end{marginnote}
Above, we first explained and operationalized what we mean by object recognition — which is operationally defined as a sensory input domain ($\sim$10 deg, $<$200 msec), and a set of behavioral capabilities within that domain (specifically referred to as "core object recognition", \citealp{dicarlo2012does}). We then emphasized that a scientifically tractable understanding of core object recognition must centrally include reproducible, image-computable models that accurately explain and predict the patterns of core object recognition behavior and the neural mechanisms underlying those behavioral capabilities. However it is not immediately clear what should comprise a neurally “mechanistic” understanding of that set of abilities.  Indeed, one  can study the mechanisms of any behavior at many different underlying levels \citep{churchland1988perspectives} and the literature in the field demonstrates myriad observations about neurons and their connections that are likely related to the mechanisms of object recognition.  This includes, for example, many reports of interesting neural functional phenomena associated with visual processing. A partial taxonomy of such reports include spatial receptive field (RF) phenomena of visuocortical neurons \citep{rust2005spatiotemporal}, surround suppression phenomena \citep{jones2001surround}, repetition suppression phenomena \citep{miller1991habituation}, various stimulus selectivity phenomena in neuron responses in key visual processing areas \citep{tsao2006cortical, pasupathy1999responses, levitt1994receptive, gallant1996neural, tanaka1996inferotemporal, logothetis1995shape}, and many, many other such seminal discoveries, far to numerous to list here. 

Given this wealth of prior work, neurally mechanistic computational models are essential to integrating these myriad phenomena into a simulation of the system -- from sensory image to multiple interacting neuronal subsystems to behavior. But what is a sufficiently mechanistic model?  A series of thought experiments is helpful here.

Suppose that one was to deliver a computable algorithm (type i system) that could take in any retinal image along with a prompt of the current task goal and it was empirically demonstrated that the behavioral “output” of that model in response to each input image precise matched — that is, could precisely and accurately \emph{predict} human perceptual report.  Would the source code of that system count as a satisfactory explanation of the mechanisms? Our guess is that, for most neuroscientists -- ourselves included -- this would not be a satisfactory mechanistic explanation. 

Now suppose that a similar algorithm was constructed to \emph{also} have a set of internal modules that each empirically behaved like  a network (\citealp{ungerleider1982two}, \textbf{Figure 2}) of specific visual brain areas (e.g., areas V1, V2, V4 etc.; \textbf{Figure 2}),  that were anatomically mapped to the hierarchical organization \citep{felleman1991distributed} of the primates’ visual cortex. For example, just like the brain, the algorithm's "V1" module was activated slight before its "V2" module, etc. Setting aside the question of its empirical accuracy, this type ii system is now at least slightly engaged on the question of mechanistic explanation.

Going further, now suppose that a type ii model of visual areas was constructed to consist of only approximations of individual simulated "neurons" in each of those areas and their connections with other model neurons in the other areas (type iii model). That is, that new overall model would be a collection of model neurons, organized in a collection of model visual regions, that work together to give rise to a computational simulation of how any image is processed by those neurons to give rise to a behavioral report.  Clearly, this type iii model is strongly engaged on the question of mechanistic explanation.  That is, unlike the type i system (above) this system is not only an algorithm -- it is also a \emph{model of the integrated set of mechanisms}.  It is a mechanistic scientific hypothesis.  

%(We note here that this is the mechanistic level of the current leading models of core object recognition that we summarize in Section 2 below.)

\begin{figure}[h]
\includegraphics[width=15cm]{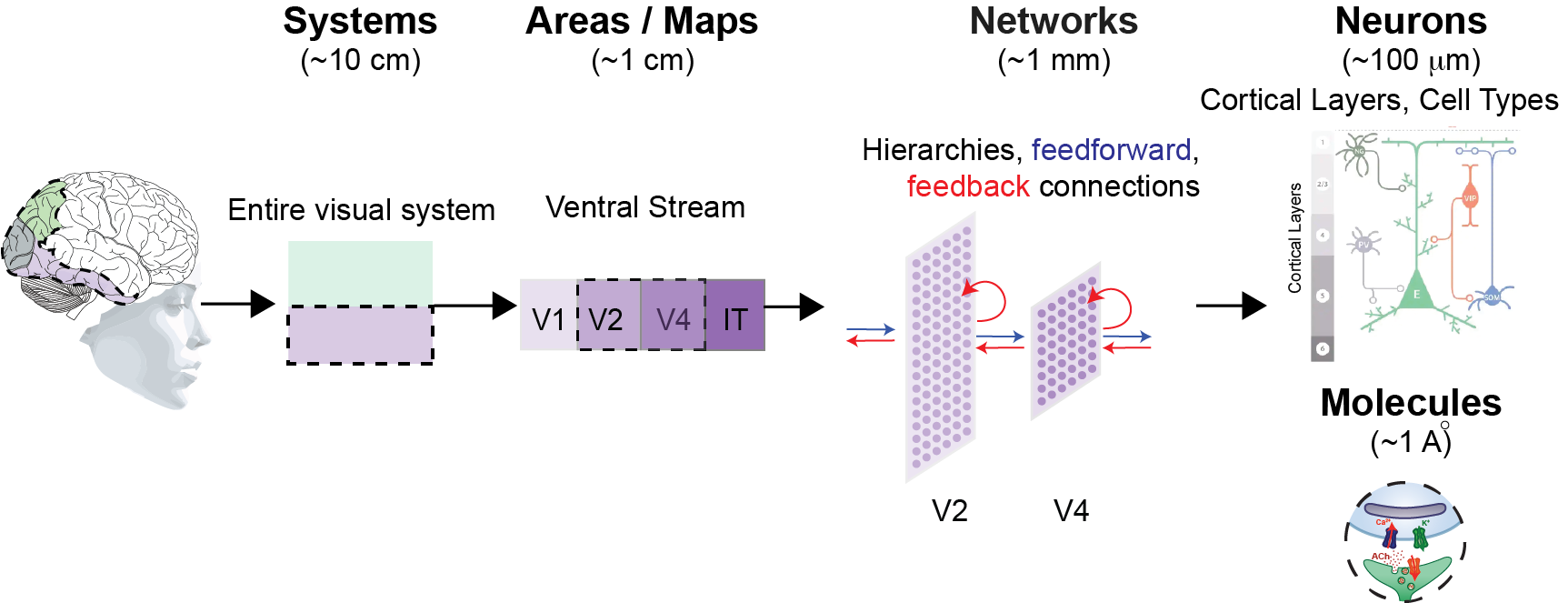}
\caption{\textbf{Constituents of a mechanistic understanding of object recognition with increasing levels of detail}. The image shows a gradual progression from systems level (the dorsal and ventral stream of the primate visual system shown here) to areas (ventral stream areas shown here) to networks (feedforward and recurrent connectivity within the extrastriate areas V2 and V4 shown here) to neurons (schematic of arrangement within a cortical layer shown here, image adapted from http://knowingneurons.com) and molecules (a single synapse shown here, sourced from https://scidraw.io/). A thorough mechanistic understanding should gradually incorporate all levels (motivated by \citealp{churchland1988perspectives}) of detail in a model.}
\label{fig2}
\end{figure}

Continuing even further, now suppose that a type iii model was constructed to \emph{also} incorporate detailed cortical layer type structures and connectivity anatomy, different morphological and genetically defined neuronal cell types and associated synaptic transmission mechanisms, and biophysically verified dendritic models; \textbf{Figure 2}; bottom row).  That new type iv  would make quantitative contact with biophysics, thus linking to agreed-upon fundamental notions of "mechanism."  In that sense, type iv models that successfully integrate all these levels would, in effect, achieve a guiding dream of our field -- to accurately and causally bridge from molecules to minds in visual object recognition.  For example, an accurate type iv model would allow us to predict precise changes in object perception that would and would not result from specific molecular interventions.

The overall point is that there is no single set of "mechanisms" of object recognition that we are after. Instead, that biological capacity -- like all cognitive capacities -- can be explained at increasing levels of mechanistic detail.  It is in this context that we next outline the state of our current mechanistic understanding, as captured in reproducible, sensory computable models. We expect that our field will increasingly develop ever more precise models that make contact with ever-finer spatial scales (see Section 4).
%which will improve a model's empirical alignment with the neural and behavioral measures, and will also unlock important new direct intervention strategies (see Section 5.).  
The current leading models (reviewed below in Section 2) are type iii explanations of mechanisms (above).

\section{SMART models of the mechanisms of core visual object recognition}

As outlined above, a critical rallying goal in understanding object recognition is the building of accurate models of the integrated set of underlying neural mechanisms and their support of object recognition behavior.  \textbf{This is an incredibly ambitious scientific goal:} the expected generalization regime is effectively infinite -- a successful hypothesis (aka model) must be accurate for \emph{any} pattern of photons that impinges on the central 10 deg of the retinae, must accurately explain any object-related perceptual judgment that can be accomplished within 200 msec of viewing time (see Section 1.1), and must ultimately explain all of the functionally-relevant neural phenomena in that same spatial and temporal window -- at least at the specified level of mechanistic resolution (See Section 1.3). 

\begin{textbox}[h]
    \section{Box 2: SMART models}

\noindent
\textbf{Sensory-computable}:  All predictions can be computed for any sensory input.  At least one behavioral report paradigm should be part of the model.  For SMART models of core visual recognition, that sensory input is the spatiotemporal pattern of photons on the central 10 deg of the retinae. The primary behavioral measures of interest are subject reports of the values of object-associated latent variables (e.g., category, identity, position, pose, etc.).

\vspace{5pt}
\noindent
\textbf{Mechanistic} and \textbf{Anatomically Referenced}:  All major model components are mapped (i.e., permanently assigned) to a part of the brain. For ventral stream SMART models, the primary brain areas of interest are the four cortical areas of the ventral stream (V1, V2, V4, and IT) along with the retina and lateral geniculate nucleus (LGN). Current mappings are limited to type ii and not type iii level of mechanisms, that treat each layer of the models as a collection of neurons from a specific brain area without specifying any level of detail about their connectivity with each other or to other brain areas (see section 1.3).

\vspace{5pt}
\noindent
\textbf{Testable}: Given the above, a model will make predictions for precisely how different (mapped) parts of the brain will respond to any given test image.   Successful predictions will support our field's belief in a particular model or set of models.  And failed predictions will reduce that belief.

\end{textbox}

Because the term "model" is used in many ways, we aim to be more precise here. In particular, we seek models that are Sensory-computable, Mechanistic, Anatomically Referenced, and Testable (referred to as SMART models; see Box 2). With this perspective, the goodness of our understanding of core object recognition (equivalently, the goodness of our current leading SMART models) should and can be primarily gauged by the accuracy with which those models explain and predict the myriad existing and future findings from all the relevant underlying brain components in the very broad regime outlined above.   

Here in Section 2, we summarize where the current leading SMART models of core object recognition came from, and the neural and behavioral observations that they have been shown to explain and predict.  In section 3, we summarize explanatory gaps that still need to be bridged with new SMART models.  In Section 4, we outline strategies to develop the next generations of SMART models.

\subsection{A sea change in neuroscience's approach to understanding the mechanisms of object recognition}

Many neuroscientists have been trained in “bottom-up” approaches where it is assumed that the study of low-level anatomical building blocks of a brain system (synapses, neurons, connectivity patterns, etc.) and the study of simplified functional phenomena (tuning functions, parameterized stimuli) can ultimately be pieced together to derive a type (iv) mechanistic model of core object recognition. As we describe below, that approach has now been turned on its head -- "top-down" integrated models that aim to achieve capabilities like object recognition are now providing the scaffold to explain and understand those myriad bottom-up measurements.

Importantly, however, some bottom-up work in primates set the foundation for that sea change.  In particular, several decades of neuroanatomical cortico-cortical tracing studies \cite{felleman1991distributed}, neuronal lesion studies \citep{phillips1988dissociation,gross1978inferior} and neural recordings studies  identified the set of cortical processing stages collectively referred to as the ventral visual stream \citep{logothetis1995shape, hung2005fast, majaj2015simple, gross1972visual, tanaka1996inferotemporal} as critical for core object recognition. The ventral stream consists of the primary visual cortical area V1, area V2, area V4, and the inferior temporal cortex (IT) (Figure 2). The input to this ventral stream starts at the retina followed by further processing at the  lateral geniculate nucleus of the thalamus, which then projects predominantly to cortical area V1, the first stage of the ventral stream. 

Exploration into the nature of neural representation (i.e. the population pattern of neural firing in response to an image) in each of these cortical areas started with the seminal findings from Hubel and Wiesel in cat primary visual cortex \citep{hubel1962receptive,hubel1968receptive} and macaques \citep{hubel1968receptive}, and extended up to the apex of the ventral stream \citep{perrett1993neurophysiology, tanaka1996inferotemporal, logothetis1995shape, tsao2006cortical, hung2005fast}. Several organizing observations have been repeatedly made in the ventral visual pathway. For instance, researchers have observed an increase in the receptive field size of neurons along the hierarchy and a corresponding delay in mean neuronal response latency.  In particular, in the central 10 deg, RF sizes progress as: $\sim$1 deg (V1), $\sim$2 deg (V2), $\sim$4 deg (V4), $\sim$10 deg (IT).  And latencies progress as: $\sim$50 msec, $\sim$60 msec, $\sim$70 msec, $\sim$90 msec, respectively  
%and IT RFs have a median size of \~10 deg and almost always include the center of gaze (latency \~ 70-120 msec)
%V1 RFs range from 0.5-2 deg (latency $\sim$40-70 msec), V2 RFs range from 1-3 deg (latency $\sim$50-80 msec), V4 RFs range from 2-6 deg (latency $\sim$60-100 msec) 
\citep{dicarlo2012does, gattass1981visual,gattass1988visuotopic, op2000spatial}. In addition, the stimuli selectivity (i.e., how narrowly tuned to a specific type of natural stimuli or stimuli features neurons are) also tends to vary across these pathways. Specifically, while V1 neurons have small RFs and are nearly optimally driven by oriented (Gabor) patterns of light patterns \citep{ringach2002orientation}, V2 neurons show preferential activations for various textures \citep{hegde2000selectivity,freeman2013functional}, V4 neurons for curvatures \citep{pasupathy1999responses}, and IT neurons for a range of semantically meaningful concepts like faces \citep{tsao2006cortical}, bodies \citep{doi:10.1146/annurev-vision-100720-113429}. Most of these observations were conducted with a limited set of hand-crafted, parametric images. In large sets of natural images, IT neurons have much more heterogeneous stimulus selectively \citep{hung2005fast, majaj2015simple}.  Implicit in many of these studies in areas V1, V2, and V4 was the observation that the selectivity properties at each stage of visual processing were approximately spatial shift invariant -- that is, different neurons had the same functional selectivity as others (e.g., a preference for rightward tilted gabors), but operating in parallel at a fully tiled set of locations across the visual field.

%Despite the myriad studies, a fully integrative, computational model comprising these brain components, that could accurately solve object-recognition tasks at human-level accuracies and explain the observed neural phenomena in the ventral stream remained elusive. 
Together, these "bottom-up" observations, along with the anatomical tracing studies, pointed to a stacked, feedforward architecture with complete sets of shift-invariant neural spatial filters at each cortical stage as the scaffold of ventral visual processing (reviewed by DiCarlo 2012).  That scaffold architecture is today known as a deep convolutional neural network (DCNN), a particular subtype of ANN models \citep{yamins2016using}. Historically, the DCNN architectural family of models descended from work as far back as \citealp{fukushima1980neocognitron}, and later work by \citep{rumelhart1986learning, lecun1995convolutional, riesenhuber1999hierarchical}. Nevertheless, despite forty years of such bottom-up work following the seminal work of Hubel and Weisel, the field had not produced models -- DCNNs or otherwise -- that could solve the hard problem of core object recognition. 

However, beginning in 2012, the field of visual neuroscience witnessed a sea change in approach.  This change began with the emergence of some artificial neural networks (ANNs) that began to rival primates in object categorization tasks.  These ANNs were architecturally inspired by the ventral stream in that they were all DCNN subtypes of ANNs, thereby incorporating evidence from the "bottom-up" approach, as outlined above.  Importantly, however, these new, high-performing DCNN models were also guided by a "top-down" behavior optimization goal -- successful assignment of each image to one of many object categories (e.g., \citealp{russakovsky2015imagenet}).  Progress toward that goal was fueled by optimization techniques that allowed the setting of the myriad network parameters that the bottom-up neuroscience functional phenomena could not determine.   \citep{krizhevsky2012imagenet, yamins2013hierarchical, yamins2014performance}. These (DCNN) ANNs turned out to have unprecedented high performance on object recognition tasks and can be considered a key breakthrough point in the evolution of SMART models. The advent of high-performant DCNNs does not trace back to a single event, but a combination of improvements around labeled image data availability, compute availability, architectural modifications, and optimization improvements. While a comprehensive history of those is beyond the scope of this review, we provide a synopsis of key milestones in object recognition in Box 2 (also see \citealp{yamins2016using}, for review). 

The key advances with respect to SMART models of primate core object recognition were demonstrated between 2012-2014.  First, by their demonstrated "behavioral" level successes, some DCNNs were quickly elevated to the current best explanations of object recognition at the mechanistic model type i level (see Section 1.3 above).  Second, the arrival and availability of these high-performant DCNN models enabled researchers to discover -- surprisingly to many -- that some of these models were among the leading hypotheses at mechanistic type ii and type iii levels (Section 1.3 above).  Most notably, it was discovered that the internal simulated "neurons" in these models were highly functionally similar to biological neurons along the ventral visual stream and  significantly better than previous models in the field (see \citealp{yamins2016using, schrimpf2018brain} for review). 
 
In this review, we focus on the many successes and current weaknesses of this top-down, achieve-behavior-first approach (aka "performance optimization approach", Yamins 2016) to building a mechanistic understanding (aka SMART models). We see this approach as important and synergistic with the more traditional bottom-up neuroscience approaches (as described in Sections 4 and 5).  But first, we review the empirical agreement of current SMART models of core object recognition with a range of neural and behavioral measurements.

\begin{textbox}[h]
    \section{Box 3: Evolution of ANN-based SMART models of ventral visual processing}
    
The perceptron model by Frank Rosenblatt \citeyear{rosenblatt1958perceptron} laid the foundation of ANNs, with refinements from Minsky and Papert \citeyear{minsky1969introduction}. Building upon the ANN image recognition system, neocognitron by Fukushima and Miyake \citeyear{fukushima1980neocognitron}, Yann LeCun's team introduced convolutional neural network ANNs (CNNs) in the 1980s. The concept of a complete set of local spatial receptive fields tiling all of the visual field, each with a similar function, is rooted in Hubel and Wiesel's work \citeyear{hubel1959receptive}, and was implemented as ``convolutions" in CNNs. The potential of CNNs for invariant form recognition was showcased with LeNet's successful handwritten digit classification \citep{lecun1989backpropagation}. By the 1990s, several hierarchical CNN models, including the notable HMAX, explained by Logothetis et al. \citeyear{logothetis1995shape} and refined by Serre and Riesenhuber \citeyear{serre2004realistic}, emerged. The mapping of such models to the ventral stream produced the first generation of SMART models of ventral visual processing.

The trajectory of CNN research reached a significant milestone with Krizhevsky et al.'s 'AlexNet' in 2012 \citep{krizhevsky2012imagenet}. Defined by their ventral-stream-like deep architecture (but optimized via non-biological supervised backpropagation), these models, beginning with AlexNet, set benchmarks in the ImageNet challenge \citep{russakovsky2015imagenet}. The large ImageNet image set and associated competition turned out to produce products (i.e., deep ANN models styled to approximate the feedforward anatomy of the ventral stream) that have revolutionized visual neuroscience. 

With high-accuracy ANN models available, inquiries arose about their relationship to neural responses in visual cortices. Yamins et al. \citeyear{yamins2013hierarchical, yamins2014performance} introduced an alternative to AlexNet, emphasizing performance optimization and architectural search. Their hierarchical modular optimization (HMO) model not only excelled in object categorization but also aligned with primate brain structures like V4 and IT.  Thus began the second generation of SMART models of ventral visual processing. Subsequent studies validated that backpropagation-trained CNNs reflected human and macaque ventral stream activities \citep{khaligh2014deep, cadieu2014deep, gucclu2015deep}. Convolutional neural networks have also become the leading models for explaining the previously well-studied responses of early visuocortical neurons \citep{cadena2019deep}, V4 neurons \citep{bashivan2019neural, pospisil2018artiphysiology}, and retinal ganglion cells \citep{mcintosh2016deep}.

\end{textbox}

% Heading 3
\subsection{Empirical tests of current SMART models}

``It doesn't matter how beautiful your theory is, it doesn't matter how smart you are. If it doesn't agree with experiment, it's wrong."  \\  
--- \hfill 										Richard Feynman\\

 SMART models of object recognition have made significant strides toward emulating human object recognition capabilities. For an up-to-date complete list of the currently leading SMART models and their empirical evaluation, we point the reader to the open science Brain-Score platform (http://brain-score.org), \citealp{schrimpf2018brain} (note that this platform refers to SMART models as "Brain models"). 

\subsubsection{Behavioral prediction tests of SMART models}
 
 Initially, as outlined above (2.1), a foremost objective of the ANN precursors of SMART models was to achieve human-level performance in terms of \emph{mean} accuracy over many categories, which has been a primary benchmark in computer vision in assessing the efficacy of these models. Remarkably, some ANNs have not only reached but in some instances surpassed this threshold of \emph{mean} accuracy \citep{dosovitskiy2020image}, at least for situations that are not substantially different from typical (but see \citealp{barbu2019objectnet}). 

 \begin{figure}[h]
\includegraphics[width=15cm]{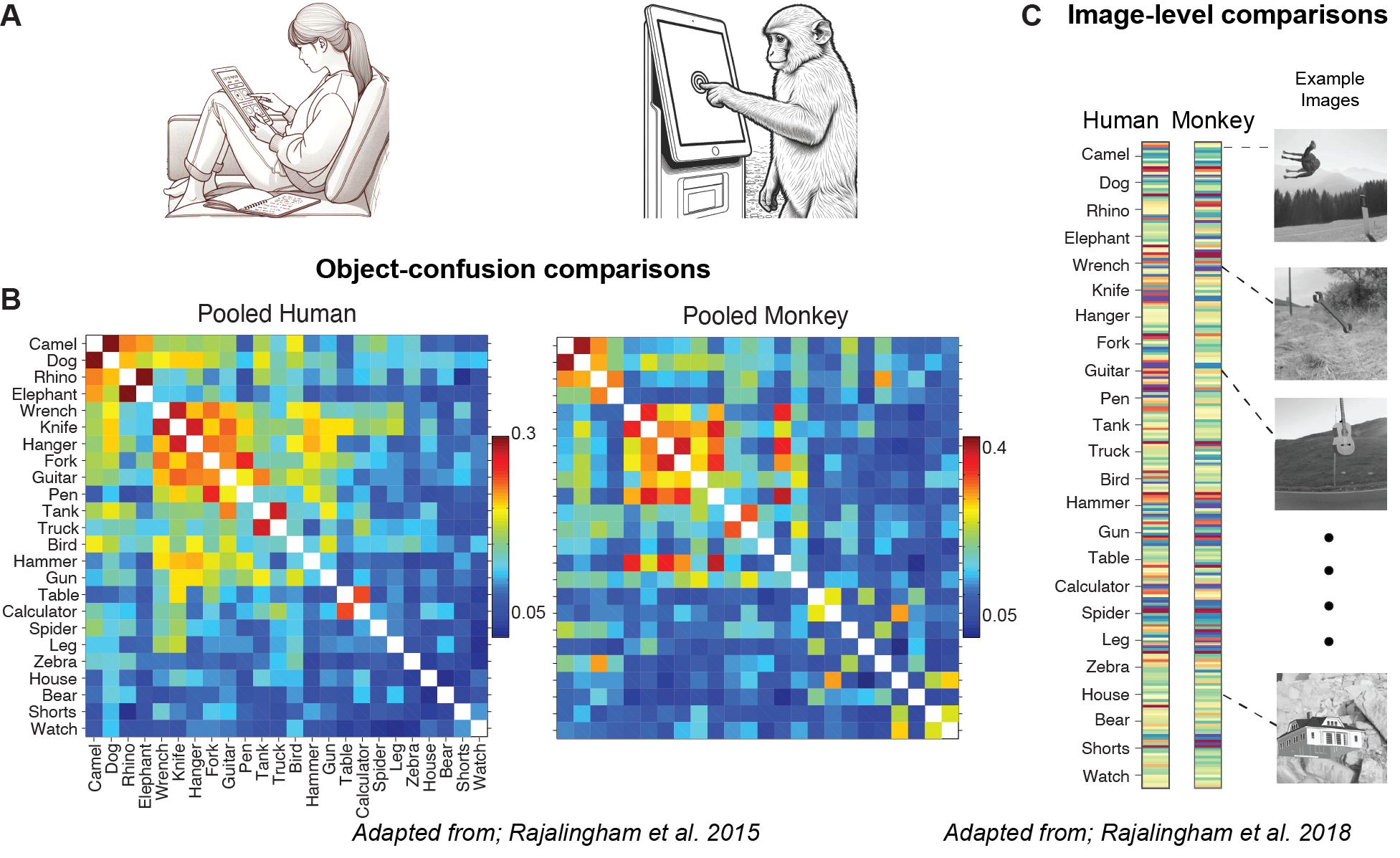}
\caption{\textbf{Comparison of object recognition behavior between monkeys and humans.} \textbf{A.} Schematic of two subjects, human and rhesus macaque, participating in a behavioral task respectively. \textbf{B.} Comparison of object confusion patterns across pooled human (left panel) and monkey (right) populations. For details, see Rajalingham et al. \citeyear{rajalingham2015comparison}. \textbf{C.} A finer grain comparison can be performed at the image level (example images shown on the right). Each value is approximately equivalent to the behavioral accuracy of determining, among a set of possible objects, which object generated that particular image (details available in \citealp{rajalingham2018large}).}
\label{fig3}
\end{figure}

 One can easily imagine a computer vision system that matches or exceeds mean human performance, but that makes mistakes that are not human-like (e.g., think of the bar code reader at your supermarket checkout).  In contrast, a fully accurate SMART model must, by definition, not just match overall average human performance, but must also make the same mistakes that humans make. Note that this is where the neuroscience/cognitive science definition of an "accurate" model (empirical alignment with the brain and its output) differs from the computer vision definition of accuracy (performance relative to ground truth).  In that regard, it is highly non-trivial that some of the high-performant DCNNs (in the computer vision sense), also turned out to have unprecedented good alignment with independently measured patterns of human object recognition behavior.  For example, for some ANN systems, objects that are difficult to discriminate are also difficult for humans to discriminate; and objects that are easy to discriminate are also easy for humans to discriminate.  In careful quantitative testing, studies report that some DCNN models are statistically indistinguishable from humans and monkeys at this level of behavioral comparison (referred to as the "consistency" of object level confusions; see \textbf{Figure 4B}, \citealp{rajalingham2015comparison}). Indeed, such strong empirical alignment observations are part of what elevates some -- but not all -- ANN systems from brain-inspired technology drivers to scientific SMART models of primate core object recognition.

The current leading SMART models and humans make surprisingly comparable errors (on object categories and individual images), suggesting a deeper, structural similarity in the way visual information is processed. This behavioral level alignment extends to more nuanced aspects of visual recognition,  and hierarchical processing of visual input, further underscoring the parallels between artificial and human visual cognition \citep{jacob2021qualitative}.  However, even the leading SMART models are not behaviorally aligned with humans in all respects (See Section 3).

\subsubsection{Neural response prediction tests of SMART models}

One quantitative way to ask if the neural mechanisms inside a candidate SMART model explain those at work in the ventral stream is to measure the functional similarity of neural representations in both of those systems.  Such comparisons can be done in several ways (See Box 4, 5), and methods and statistics around such comparisons are an active area of research \citep{yamins2014performance,kriegeskorte2008representational,schrimpf2020integrative,schrimpf2018brain}.   At their core, all of these empirical tests ask about the ability of the simulated neuronal population in SMART model area X (e.g. SMART model area "V4") to predict the neuronal population in that same ventral stream area.  The notion of "prediction" here refers to the testing of images that were never used to estimate any of the SMART model's internal parameters and were never used to estimate the model-to-brain mapping parameters (that is, which simulated neuron(s) in the model correspond to the biological neuron(s) of interest).  

Before describing some of those results, we note that, unlike SMART models, ANN or DCNN vision systems that do not have mapping commitments to brain areas cannot be tested in this way. That lack of commitment does not impugn the potential utility of those systems in other venues. Instead, it simply reflects a lack of engagement on the question of neural mechanism; see Section 1.3. 

 %In Box 3, we explain a linear regression based model evaluation method to assess this, so called. "neural predictivity". There are many other ways to compare representational spaces across models and neurophysiological data. In Box 3, we explain another common method, of representational similarity analysis (RSA).

%A number of studies have used neural predictive and representational comparison methods to evaluate alternative SMART models at all levels of the ventral stream.  

Beginning in 2013, it was discovered that the responses of their internal components -- artificial "neurons" within each of the model "areas" (aka model layers) -- often strongly align with the responses of their biological mapped counterparts \cite{yamins2013hierarchical,yamins2014performance,cadieu2014deep}. Those, and many later studied showed that current SMART models can predict $\sim$50 \% of the explainable neural response variance. This was significantly better than models from a decade ago \citep{riesenhuber1999hierarchical, serre2004realistic}, but still less than perfect (performing below the noise ceiling as estimated per neuron). It is important to note that how well (i.e., the noise ceiling) a model neuron should predict an IT neuron recorded from a randomly sampled monkey depends on many prior assumptions (are we building a model of that specific monkey? or an archetypal monkey? etc.), and is a matter on ongoing research. 
%[Say something about ceilings unclear?]  

\begin{figure}[h]
\includegraphics[width=12.5cm]{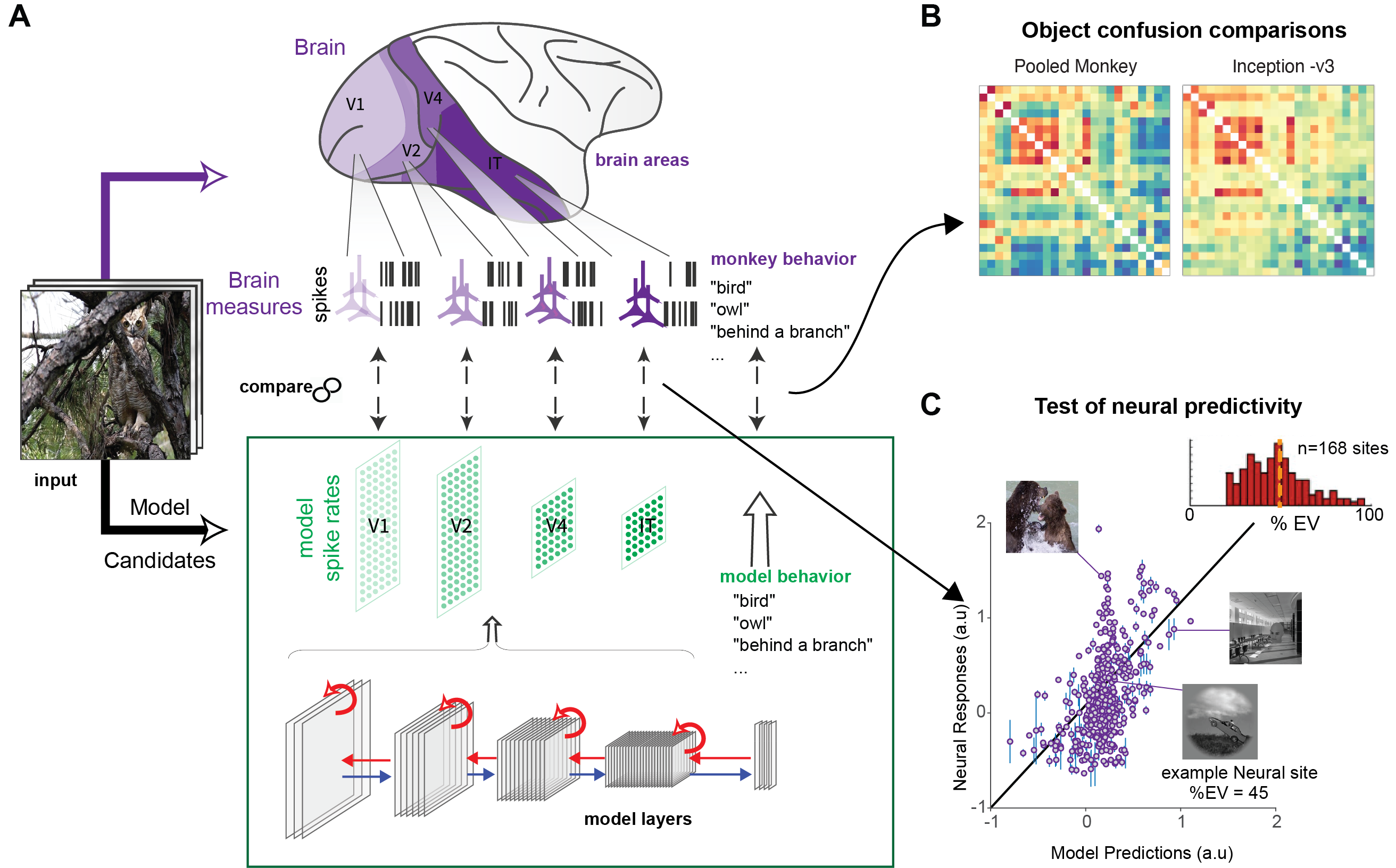}
\caption{\textbf{Evaluating the alignment of SMART models and primate behavior and neural responses.} \textbf{A.} Current SMART models (See box 2) are derived as brain-mapped, image-computable deep CNNs (see text).  The predictions of SMART models -- for a proposed experiment -- are obtained simply by performing the same experiment on the model. For instance, by presenting a planned set of images as the input to the model and ``recording" the responses of individual model neurons from a model brain area (e.g., IT) or by recording its behavioral responses. To assess the empirical alignment with biology, these predictions are compared with the results of that experiment, scored with a quantitative \emph{metric}.  Each such alignment test is referred to as a \emph{benchmark}. When multiple benchmarks are performed on \emph{one} SMART model, this is referred to as \emph{integrative benchmarking} \citep{schrimpf2018brain, schrimpf2020integrative}. Here we illustrate just one behavioral comparison (B) and just one neural comparison (C).  \textbf{B.} Comparison of the monkey behavioral object-level confusion patterns and an inception-v3-derived SMART model \citep{szegedy2016rethinking}.  The metric of alignment here is the correlation over all the values in the two matrices (see \citealp{rajalingham2018large}). \textbf{C.} Image-level neural response predictivity test (here, for one study of IT cortex).  Scatter shows results from one example IT neural site and the SMART-model predicted responses of this site (to do this, the model must be "mapped" at the level of single units, see Box 4 for how that is done). Each dot is the model predicted response (x) vs. the actual neural response (y; mean rate in a time window, averaged over repetitions of that image). The elemental alignment metric here is the fraction of image-response variance accurately predicted (EV), corrected for irreducible noise.  The overall alignment metric is the median EV overall recorded neural sites in the dataset (see inset histogram) \citealp{yamins2014performance}.}
\label{fig4}
\end{figure}

Over the past decade, many studies in the ventral stream have either explicitly or implicitly replicated this core neural finding. For example, at the spiking neural level, recent SMART models: predicted V1 responses to natural images with unprecedented accuracy \citep{cadena2019deep, dapello2020simulating}, predicted specific types of shape tuning in V4 neural responses \cite{pospisil2018artiphysiology}, and were reported to be the best predictor of anterior IT face-patch response (AM, anterior medial) \citep{chang2021explaining}. Other  studies have used SMART models to predict functional aspects of ventral stream neural representations as assessed by fMRI \citep{khaligh2014deep, ratan2021computational, cichy2016comparison, gucclu2015deep,agrawal2014pixels}, ECoG \citep{grossman2019convergent}, and MEG \citep{cichy2016comparison}. 

Given the diversity of stimuli and methods, it is still difficult to tell if there is a trend for some areas of the ventral stream to be better explained than others.  This is compounded by the fact that different areas have different functional dimensionality (in the models and likely in the biology as well), which makes such comparisons dependent on the comparison metrics.  To our knowledge, the best summary of the current state of SMART models of the ventral stream and its supported behavior is tracked on the open science Brain-Score platform (http://brain-score.org).  This platform is still far from perfect, but it is better than no tracking at all, and it continues to improve in its functionality and number of neural and behavioral benchmarks.  Inspection of the Brain-Score benchmarks suggests that a large amount of neural functional response variance is currently captured by leading SMART models, but that no model is yet fully accurate, even among the limited set of neural benchmarks that are available.  

Taken together, what all of these studies imply is that the image-driven functional response profiles of individual neurons along the ventral stream are surprisingly similar to the functional profiles of their individual "digital twin" SMART model neurons.  The representational tests imply that the population distributions of the different functional types of neurons are approximately matched.  Indeed, the leading SMART models of the ventral stream are referred to as the leading models in part because they do very well on these neural functional comparisons -- far better than earlier models.

\begin{textbox}
   \section{ Box 4: Testing the neural alignment of SMART models}

\vspace{-5 pt}
\subsection{Neural response measurements:}

\noindent
\textbf{Model:}  Measurements of the responses of SMART model neurons are made by presenting a set of test images and "recording" the scalar activation values of all neurons in the to-be-predicted model area. In AI, this is referred to as, "extracting features" from a specific layer, like the 'fc7' layer of AlexNet.
\vspace{5 pt}

\noindent
\textbf{Brain:} Experimental recordings of the responses of individual neural sites in a to-be-predicted brain area are made by presenting the same (above) set of test images (e.g. typically for 100 or 200 ms duration each). To determine each site's response, spikes are counted in a latency-adjusted time window (e.g., 70-170 ms post image onset in \citealp{yamins2014performance}), averaging over all repeat presentations of each test image (typically \>20). Because neural responses are dynamic, SMART models have also been usefully compared at finer temporal resolution (e.g. \cite{kar2019evidence}).

\subsection{Mapping SMART model neurons to biological neural units:} Most current SMART models -- when downloaded for testing -- are only anatomically-referenced at the level of brain areas (see Box 2). To make neural predictions at finer spatial grain (e.g., the response of a particular recorded single neuron or a particular measured single fMRI voxel), the SMART model ``neurons" must first be ``mapped" to that finer grain.  There are several methods to do this, each with pros and cons \citep{yamins2013hierarchical, kar2019evidence, arend2018single, klindt2017neural}, but all mapping methods assume a linear relationship between model neurons and biological neurons. That mapping is determined via responses to a set of ``mapping" images. The initial \citealp{yamins2014performance}, still a commonly used mapping method, is predicting each biological neural site as a regression on a set of SMART model "neurons", where the regression weights are determined using the mapping responses. Different regression regularization choices correspond to different mapping methods. Once the mapping is determined, it is frozen, and model predictions are then evaluated (below) using \emph{new} images.

\subsection{Metrics to assess the goodness of model predictions:}

%\textbf{Training and Prediction:}\\
%\noindent
%Step 1: Map model units to individual neurons: Use a suitable train/test split of images and estimate regression weights with a regularized regression method.\\ 
%\noindent
%Step 2: Generate the model prediction using the regression weights (from Step 1) and model activation features for the held-out test images.

%\textbf{Evaluation Metrics}\\
\noindent
\textbf{Neural Predictivity:} A measure of how well the model's predicted neural responses match the actual measured responses.  Typically in units of explained variance R${^2}$,  corrected for non-reproducible variance in the measured variables (i.e. "noise" estimated from repeat tests of the same images).

\vspace{5 pt}

\noindent
\textbf{Representational Similarity Analysis (RSA):} Pioneered by Kriegeskorte et al. (\citeyear{kriegeskorte2008representational}) (see \cite{nili2014toolbox} for details) For n test images, a 'n x n' symmetric matrix is constructed in which the element in row 'i' and column 'j' indicates the distance (e.g. Euclidean) between the neural population activity patterns corresponding to image i and image j. One matrix is constructed for the neural measurements from a brain area (e.g. fMRI voxels) and another for the corresponding SMART model neural population (in response to the same images). The two matrices are quantitatively compared (e.g., by Pearson correlation).

\vspace{5 pt}
\noindent
\textbf{Centered Kernel Alignment (CKA):}  Motivated by the idea that a similarity measure should be invariant to rotation and isotropic scaling, but not to all linear transformations, Kornblith et al. \citeyear{kornblith2019similarity} proposed CKA as another way to gauge the similarity of model and brain population representations.

\end{textbox}

%\begin{textbox}
%\section{Box 5: Representational Similarity Analysis (RSA)}Representational Similarity Analysis (RSA), pioneered by Kriegeskorte et al. (\citeyear{kriegeskorte2008representational}) is a widely recognized approach. The specifics of this method can be found in Nili et al. (\citeyear{nili2014toolbox}). In practice, it involves constructing a square symmetric matrix, where both the horizontal and vertical indices are determined by the stimuli, and organized in identical order. The elements, located at row 'i' and column 'j', indicate the difference (or distances in multivariate response space) between the activity patterns corresponding to two distinct stimuli, such as images. By design, diagonal entries, which compare identical stimuli, have a value of 0. For any entity or measurement category (like human fMRI BOLD signals, CNN model activations, and so on) that shares a stimulus set, these matrices can be formulated. Once constructed, these matrices offer a means to compare and evaluate the similarity between different datasets. Of note, Kornblith et al. \citeyear{kornblith2019similarity} suggest that a similarity measure should maintain invariance to properties like orthogonal transformations and isotropic scaling, but not to invertible linear transformations. Based on this, they proposed centered kernel alignment (CKA), as another way to gauge the similarity between two representations.
%\end{textbox}

\subsubsection{Neural control tests of SMART models}

“All models are wrong, but some are useful”, an aphorism attributed to the statistician George Box, also applies to models of object recognition. Recently, the value of the SMART models has been augmented by goal-directed stimulus synthesis of images. For instance, Bashivan, Kar, and DiCarlo (\citeyear{bashivan2019neural})  demonstrated (\textbf{Figure 5A}) that by using a SMART model that included visual are V4, they could generate synthetic stimuli that drive specific, experimenter-chosen V4 neurons to response levels beyond what could be achieved by the previously known “preferred-stimuli” for the region. They also showed that this approach could be used to target the entire sub-populations of recorded neurons -- demonstrating at least partial ability to independently set each of neurons at a desired activation state.  These tests have been referred to as neural control tests because the goal is to drive/set the neural activity level or levels to a particular, experimenter-chosen state.  A critical observation from that study was the high correlation between the accuracy of the model predictions over naturalist images and the quality of neural control that could be obtained, suggesting that the neural prediction measure (section 2.4) are reasonable proxy measure of the more applied goal of neural control.  Related experiments have been carried our in other brain areas.  For instance, Ponce et al. \citep{ponce2019evolving} demonstrated that they can synthesize “super-stimuli” for inferior temporal (IT) neurons that drive the activity of these cells beyond their usual response range (\textbf{Figure 5B}). In fact, their results challenge the common terminology in the field given that the super-stimuli for a classical “face-selective” neuron does not resemble a typical primate face — paving the way for a new set of model-based “intuitions” for how to think about neuronal encoding spaces. Similar approaches have also been implemented in the rodent neuroscientific community \citep{walker2019inception}

\begin{figure}[h]
\includegraphics[width=15cm]{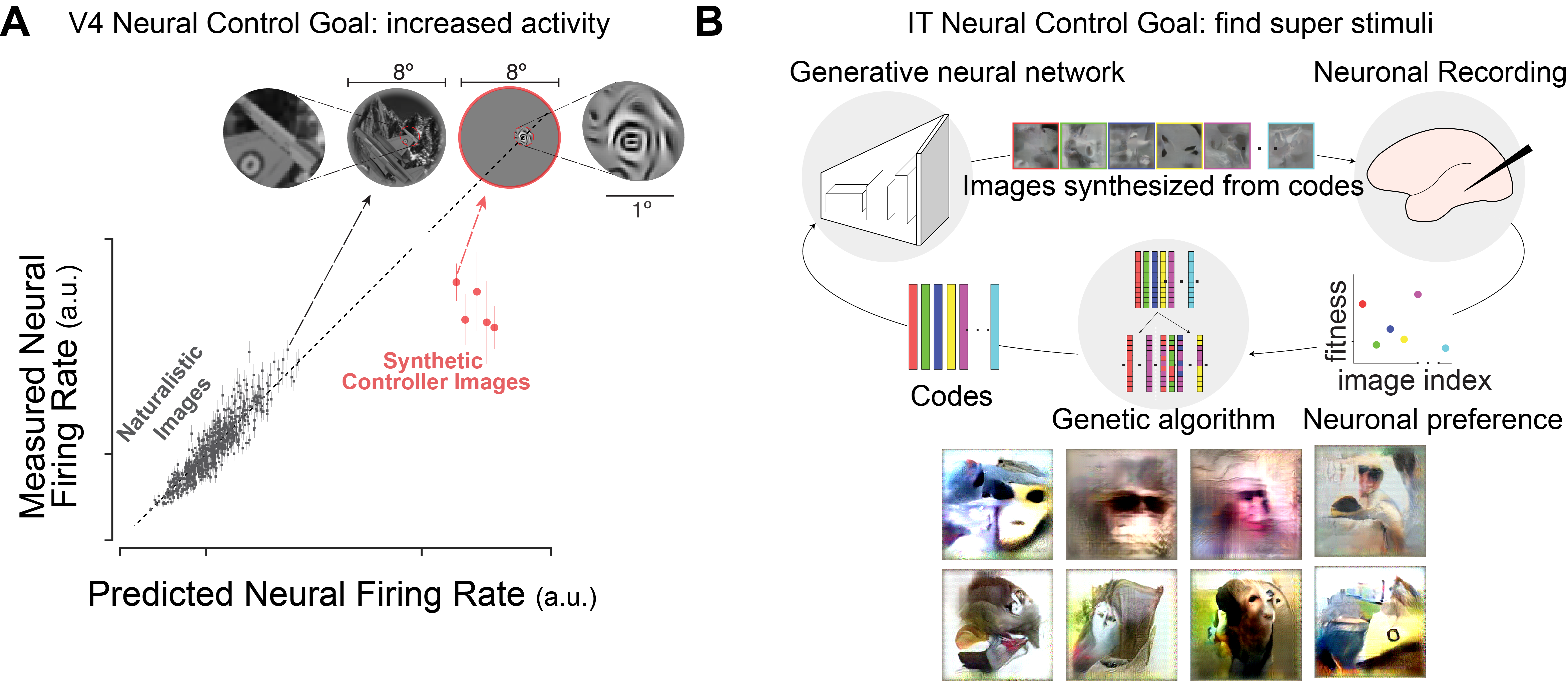}
\caption{\textbf{Examples of neural control in mid-level area V4, and in IT cortex.} \textbf{A.} Model-guided generation of synthetic images was shown to increase neural firing rate beyond the range observed in a large set of naturalistic images. For details see Bashivan et al. \citeyear{bashivan2019neural}. \textbf{B.} Schematic of the generative evolution method, XDream \citep{xiao2020xdream} in which a deep generative adversarial network was used to synthesize images presented to monkeys. Neuronal responses were used to rank and optimize the image codes using a non-gradient-based optimization algorithm, here illustrated for a genetic algorithm. The bottom panels show examples of images evolved for face neurons (top row) and non-face neurons (bottom row). For details, see Ponce et al. \citeyear{ponce2019evolving}.}
\label{fig5}
\end{figure}

\subsection{Future tests of SMART models: Direct neural perturbations}

Tests of SMART models should not be confined merely to behavioral and neural recordings where the primary causal perturbation tool is the pattern of photons on the eyes (reviewed above).  This method is a powerful, yet indirect causal manipulation of the neurons. There is a compelling case for branching out into more direct neural perturbation experiments where, in the ideal case, the experimenter injected energy could be precisely targeted first to \emph{only} the neurons under study. Such experiments, encompassing techniques like optogenetic, chemogenetic, and electrical interventions, have historically been instrumental in more strongly addressing the mechanistic role of modeled brain components in supporting behavior. Yet, despite their significance, the insights derived from these studies often fall short due to the limited and somewhat arbitrary control provided by today's direct neural perturbation tools available in primates, as observed by Jazayeri and Afraz \citeyear{jazayeri2017navigating} and Wolff and Olveczky \citeyear{wolff2018promise}. Consequently, the conclusions of such work can often only reinforce previously established  models derived from more indirect methods, leaving more definitive causal hypotheses unvalidated.

However, a silver lining emerges with the noticeable congruence between brain tissue and SMART models. This alignment offers an unprecedented opportunity to harness causal perturbation techniques for evaluating and distinguishing between alternative SMART models. Advanced Artificial Neural Network (ANN) models, for instance, empower us to assess if digital perturbations within these models can mirror the outcomes observed from real-life, in-vivo perturbations. This trajectory holds immense promise, not just for enriching our understanding of brain perturbation experiments but also for paving the way toward crafting enhanced strategies in the domain of visual prosthesis.

\section{Known misalignments between brains and current SMART models}

Despite the surprising empirical successes of the current family of SMART models described above in section 2, there is still misalignment between these models and the primate brain at both the neural and behavioral levels (the glass is not full!). Below, in section 3, we review some known misalignments of the current ANN-based SMART models with primate neurobehavioral data. Here we only discuss examples of neural and behavioral functional phenomena that current SMART models \emph{do} aim to predict, yet fail. In section 4, we discuss how next-generation SMART model could aim to predict even finer-scale phenomena.

%We do so without suggesting model changes that might remedy those misalignments, leaving discussion of new SMART models until Section 4.   

%In this section we compile examples of [neural and behavioral phenomena] [cases]  that the list of cases that emphasize the only partial success of the current SMART models in explaining neurobehavioral phenomena that 

other related neurobehavioral phenomena. 

\subsection{Behavioral prediction failures}

Recently optimized ANNs solve object recognition tasks at unprecedented mean accuracies \citep{he2016deep}. However, as of a few years ago, no ANN exhibited patterns of successes and errors across images that fully aligned (\textbf{Figure 6A}) with human patterns measured over the same images \citep{rajalingham2018large}. More targeted looks into these misalignments have revealed specific shortcomings of ANNs that make them incomplete models of human behavior.  We discuss the more prominent of these targeted analyses below. 

First, Geirhos et al. \citeyear{geirhos2018generalisation}, observed that some leading ANNs at that time (e.g., VGG-19,  ResNet-152) were less robust (compared to humans) to the addition of Gaussian noise to images during object categorization (\textbf{Figure 6B}). Interestingly,  Geirhos et al. \citeyear{geirhos2018imagenet} also discovered that those ANNs relied more on the texture of the objects compared to their shapes (\textbf{Figure 6C}), while humans typically rely more on object shape in comparable tests. 

A second area is the behavioral susceptibility of ANNs to so-called "adversarial attacks" \citep{goodfellow2014explaining}. In brief, given the full observability of all ANNs, optimization methods have been used to search through high dimensional pixel space and successfully find small amplitude image perturbations that strongly change the behavioral output of the ANN (e.g. changing the output from "dog" to "church").  The (Euclidean) pixel amplitude of these "attacks" is typically less than a few percent of the distance between arbitrary natural images, and it was demonstrated that human behavior is largely (but not completely, \citealp{elsayed2018adversarial}) insensitive to the same changes.   This suggests a potential mismatch of those early SMART models with human vision.  At the time of this writing, tests on newer SMART models, which also have higher neural alignment \citep{schrimpf2018brain, guo2022adversarially}, reveal that human perception can be surprisingly and strongly modified by similarly small amplitude image perturbations \citep{gaziv2023robustified}.  And, when properly compared, these current leading SMART models have far less behavioral misalignment with human perception than the original adversarial work \citep{gaziv2023robustified}. But a gap nonetheless remains. 

Third, other phenomena of visual perception are thought to not be well-predicted by current SMART models.  For example, local vs. global shape processing phenomena,  dependence of object classification on object part relationships, "filling in" illusory phenomena, and "uncrowding" phenomena \citep{bowers2022deep}.  However, these putative behavioral gaps have not been formally tested. Scientific caution is warranted here, as SMART models continue to unexpectedly predict things that were not part of their explicit design and thus one might not expect them to predict \citep{ngo2023clip, fan2023challenging}.  These are now active areas of model-to-human empirical comparison studies.  

\subsection{Neural prediction failures}

As outlined in section 2, current SMART models are surprisingly accurate at predicting neural responses in areas across the ventral stream, even at the single neuron level. However, even the current best SMART models only predict 50-60 $\%$ of the explainable variance in the neural responses in V4 and IT (\textbf{Figure 6D}; for most updated statistics refer to Brain-Score). This exposes an apparent “explanatory gap” that still remains to be bridged. A more targeted investigation of these misalignments reveals that current SMART models are not fully accurate models of ventral visual processing, even at their currently intended mechanistic level (type iii, see Section 1.3 and Section 4).  We discuss a few of these targeted analyses below. 

First, when looking specifically into the functional subtypes of neurons, like face-selective neurons of the IT cortex, \citealp{chang2021explaining}
reported that CORnet-S (a leading SMART model) only predicts $\sim$50\% of the explainable neural variance. Also, the layers of the current SMART models and the brain areas of the primate ventral stream are not strictly hierarchically aligned, necessitating more careful investigation of how signals across these areas are integrated over time and how the models could explicitly implement those computations \citep{sexton2022reassessing} 

Second, the neural dynamics of current SMART models are clearly not in line with that of the ventral stream.  For example,  Kar et al. \citeyear{kar2019evidence}, working at the spiking neural level,  recently found that, while feedforward ANNs do quite well at predicting the IT population pattern in the early phase of the neural responses (90-120 ms after image onset), they are poor to moderate at predicting the late-phase of neural population pattern (150-200 ms after image onset). As shown in \textbf{Figure 6E}, this difference increases for images (labeled as ``challenge" images in Kar et al. 2019) where primates outperform baseline ANN models such as AlexNet). This observation is consistent with the lack of recurrent connections in these ANNs that other results suggest are critical in shaping the late phase of the IT response \cite{kar2021fast}.  

Third, Bashivan et al. \citeyear{bashivan2019neural} observed that current SMART models at that time did more poorly at predicting V4 neural responses to strongly “out-of-domain” images, a finding also demonstrated for IT responses \citep{ponce2019evolving}. Bashivan et al. \citeyear{bashivan2019neural} also observed that the “one-hot population” control paradigm (where the objective was to design images that only activate one neural site while not activating all the other measured V4 sites; shown in \textbf{Figure 6F}, top panel), could not be perfectly executed (as shown in \textbf{Figure 6F}, bottom  panel). 

% Third, ...

These observations collectively point towards the inherent limitations of current SMART models, even at their currently intended mechanistic level, and emphasize the pressing need for iterative advancements in our modeling approaches to encompass the intricate nuances of the primate ventral stream neural machinery.

\begin{figure}[h]
\includegraphics[width=12.5cm]{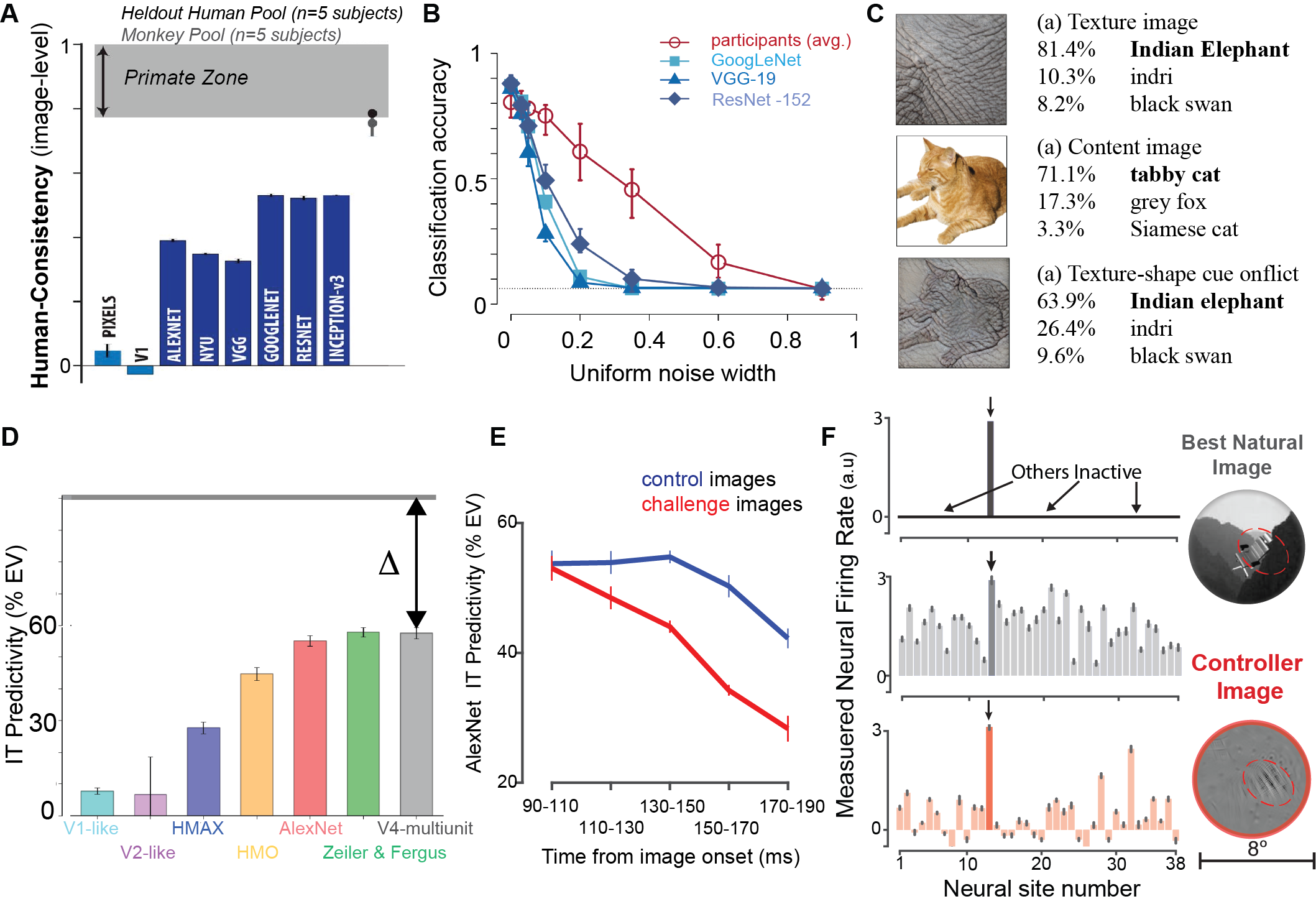}
\caption{\textbf{Explanatory gaps in behavioral and neural predictions.} \textbf{A.} Consistency of image-level accuracies during object discrimination tasks between humans and several tested models. (see Rajalingham et al. \citeyear{rajalingham2018large}. \textbf{B.} Changes in classification accuracy with increasing levels of image noise across VGG-19, GoogleLeNet, ResNet-152, and human participants demonstrates greater noise robustness in human vision \citep{geirhos2018imagenet} \textbf{C.} Classification of ResNet-50 of a) elephant skin (only texture cues), b) a normal image of a cat (consistent shape and texture), c) an image with texture-shape cue conflict (shape of a cat, texture of an elephant) shows the so-called "texture-bias" in leading SMART models at that time \citep{geirhos2018imagenet}. \textbf{D.} Neural response predictivity in IT cortex for SMART models at that time. The $\Delta$ denotes the explanatory gap. For the most up-to-date measures, check http://brain-score.org \textbf{E.} Relative to the early IT response (90-120 ms after stimulus onset), feedforward SMART models are poor at predicting the late IT population response (150-200 ms).  This gap is particularly prominent for test images for which monkeys behaviorally outperform the models (``challenge images"; red line; AlexNet shown here) compared to images where models and humans have similar performance (``control images; blue line) For  details, see \citealp{kar2019evidence}. \textbf{F.} The control objective of the “One-hot” population (OHP) is to selectively increase the activity in one neural site while keeping responses of all recorded neural sites close to zero (top row). The middle row shows the naturalistic image that most closely accomplishes this objective. The bottom row shows a SMART-model-driven synthetic controller image that performs much better. But it still does not fully achieve the OHP objective \citep{bashivan2019neural}.}
\label{fig6}
\end{figure}

\section{Where will the next generation of SMART models come from?}

As outlined above, over the last decade, the field has made major strides in developing reasonably accurate approximations of the integrated set of neural mechanisms that support visual object recognition near the center of gaze (see Section 2).  However, the field’s current best models of those mechanisms (referred to here as SMART models) still have limitations.  First, the current best models still do not yet account for (i.e. predict and control) 100$\%$ of the neural and behavioral functional measures of “core” ventral visual processing that they already aim to account for (see Section 3). Secondly, it is still unclear if simple variants of the current models can or cannot account for visual processing and visual behaviors beyond the central ten degrees and beyond what is achieved in the first $\sim200$ msec of visual processing in a default attentional state. Lastly, current SMART models do not yet map to — and thus cannot yet account for — the potentially different functions of different cortical layers, anatomical recurrences, and diverse neuronal cell types (including cells with different morphologies and genetic profiles). 

We believe that these three challenges are tightly interrelated. Specifically, performance gains in visual object recognition led to more accurate models of the neural mechanisms (see Fig. 1 in \citealp{yamins2014performance}). Therefore, we anticipate that the next-generation SMART models that achieve performance gains in other visual intelligence capabilities that critically rely on neural activity beyond the first 200 msec will lead to even more accurate predictions of the neurobehavioral phenomena associated with core visual processing.  They will likely also align better with other functional measures (e.g. IT dynamics; similar to demonstrated in \citealp{kar2019evidence}) and other brain systems (e.g. prefrontal and dorsal visual stream areas).  Furthermore, as form often shapes function, next-generation SMART models of the ventral visual stream that appropriately incorporate finer-grain biological form (e.g., recurrent anatomy, cortical circuit motifs, cell types, etc.) will likely lead to more accurate models of those neural and behavioral functional phenomena.

As such, we advocate for three forward-looking research directions: 1) Directly improving upon current SMART models of the ventral visual stream and its support of visual object recognition with more targeted experiments, 2) Building and evaluating alternative SMART models of primate/human visual intelligence capabilities beyond visual object recognition, and 3) Building and evaluating alternative, cellular-level implementations of SMART models of the ventral visual stream. While we forward the production of accurate SMART models as the rallying goal, those advances will not be possible without even tighter iteration with neural and behavioral experiments. 

The good news is that some visual neuroscientists and cognitive neuroscientists are actively engaged in direction 1, some cognitive scientists and computer scientists are already actively engaged in direction 2, and some cellular and systems neuroscientists are already actively engaged in direction 3. What is most exciting is that the current SMART models of visual object recognition (outlined above) and the open science platform for evaluating those models (Brain-Score) are now acting as a community scaffold to bridge these three fields in the domain of visual intelligence/perception.  As the modeling and experimental evaluation scaffold expand together (1-3, above), the opportunities that will unlock are myriad.  For example, once these next-generation SMART models are sufficiently accurate, this understanding will allow the field to unlock the possibility of, for example, precise predictions of how the intervention tools of pharmacology and potentially genetics (which are reasonably understood at the cell type level) will or will not influence visual intelligence capabilities at the cognitive level.

Next, we highlight some ongoing and forward-looking activities and ideas in each of these three interrelated research directions.

\subsection{Directly improving upon current SMART models of ventral processing}

A basic recipe to search among alternative SMART models of ventral visual processing has been discussed by Yamins and DiCarlo \citeyear{yamins2016using} and formalized in the deep learning framework by Richards et al. \citeyear{richards2019deep}. In brief, the key components to be explored are model architecture (the functional building blocks of the model), behavioral objective (the computational goal of the model, e.g., object categorization), and the learning rules (how the model is optimized with its set architecture to accomplish the behavioral goal). 

Ongoing work is aimed at exploring these components. The primary goal of many of these studies thus far has not been to improve the empirical accuracy of current SMART models on adult functional measures (Section 2). Instead, these studies: 1) aim to find minimal conditions that might give rise to SMART models with similar empirical accuracy as ANNs discovered by computer vision via bio-implausible optimization methods, and 2) extend the set of adult phenomena that SMART models accurately account for.  This includes, for example, efforts to explore more plausible evolutionary selection mechanisms \citep{geiger2021wiring}, more biologically plausible learning rules that might unlock new hypotheses about postnatal visual development \citep{zhuang2021unsupervised}, more ecologically-relevant experience histories \citep{barbu2019objectnet,mehrer2021ecologically}, and/or mechanisms that can also explain the topographic organization of the ventral stream \citep{lee2020topographic, margalit2023unifying, dobs2022brain, doshi2023cortical}. These important normative activities each seek to develop new variants of SMART models that can explain not only how the ventral stream works the way it does, but why it works the way that it does, and how it got to be that way. 

In addition to this ongoing normative research, recent work has also been aimed at using adult functional data to \emph{directly} guide the building of more accurate SMART models of those types of data. For example, Dapello et al. \citeyear{dapello2022aligning} directly used neural recordings from the IT cortex to regularize the training of ANNs (alongside ImageNet categorization loss), leading to more human-aligned ANNs that better predicted neural responses on new monkey subjects and images, behavioral patterns and also became more adversarially robust. In a similar approach, but with behavior, Fel et al. \citeyear{fel2022harmonizing} developed a ``neural harmonizer," training method that aligns ANNs with human visual strategies while also enhancing categorization accuracy on new images.  These "direct model optimization" approaches may not sufficient on its own in the near term, due to current data limits.  Nevertheless, as experiments are beginning to produce ever-larger volumes of functional primate and human data, we suspect that this strategy will also be an important part of discovering next-generation SMART models of the ventral stream.

\subsection{Evaluating alternative SMART models of ventral visual processing}

In addition to the approaches to \emph{build} new models above, it is just as important to highlight the importance of methods to more reproducibly and efficiently \emph{test} and \emph{adjudicate} among alternative models. These include for example, benchmarking platforms to collect and maintain all past tests \citep{schrimpf2018brain}, methods to pit SMART models against each other to discover ‘controversial’ stimuli on which their predictions most disagree \citep{golan2020controversial}, and methods to interpret the results of such tests \citep{canatar2023spectral}.  For example, the Brain-Score benchmarking platform (http://www.brain-score.org) aids various labs in evaluating the impact of different modeling approaches in producing more empirically accurate SMART models of ventral visual processing and its supported behavioral capabilities. Together, such methods unlock more efficient scientific hypothesis adjudication, a core community activity of any reproducible science.

\subsection{Building SMART models of visual intelligence beyond object recognition}

Human visual intelligence is not just object recognition.  And it derives from the \emph{entire} visual system, not just the ventral stream.  Indeed, even the ventral stream's function in the brain is not limited to core object recognition. It is involved in other visual tasks, including scene understanding, expression estimation, and more. If we limit our models only to object recognition, we certainly will not be able to understand the mechanisms of all of visual intelligence.  And even if one only cares about the ventral stream, not considering its broader role means that we would likely miss out on capturing mechanisms that it contains for supporting visual intelligence, but that are not necessary for supporting core object recognition.  
%Modeling how ventral stream networks evolved as a part of a larger circuitry with varied optimization goals might be critical in developing a comprehensive digital twin of the primate ventral stream and for the broader range of visual tasks that it supports. 

The ventral visual stream, often termed the ``what pathway," has been traditionally associated with object recognition and form representation, while its counterpart, the dorsal stream, often called the ``where pathway," has been associated with representing spatial location, motion, and guiding actions like grasping. Many recent studies have shown that the functional specializations of these pathways are more complex and often overlapping \citep{de2011usefulness}. In addition, recent developments in computer vision also facilitate incorporating other behavioral tasks like object detection \citep{zhao2019object}, and monocular depth prediction \citep{zhao2020monocular} etc. Beyond recognizing individual objects, our brain processes entire scenes, recognizing actions and interactions of various agents \citep{mcmahon2023seeing}, understanding contexts \citep{zhang2020putting}, and making predictions based on the environment and the physics of the world \citep{bear2021physion}. Therefore, SMART models can be developed to understand how these two pathways interact and integrate visual information. 

More broadly, to truly understand and model human visual intelligence, we must venture beyond just object recognition and delve into the myriad of other tasks our visual system performs. This is likely to involve not only what are now mainstream ANN optimization methods, but also to be accelerated by modeling methods that begin with symbolic structure, can generate alternative internal predictions at some level of representation \citep{lake2015human}, and can explicitly manage probabilistic inference in a manner that can scale \citep{gothoskar20213dp3}.  Indeed the field is now seeing a fusion of such approaches with ANN optimization methods and, when neurally mapped, this will produce new SMART models that will need to be experimentally adjudicated.  

\subsection{Building and evaluating alternative, cellular-level implementations of SMART models}

Many aspects of the known primate brain circuit architecture are currently not explicitly mapped onto the current SMART models. While it is possible that functional approximations of such motifs are already present in these models in some form, an explicit mapping is definitely missing rendering these models less interpretable \citep{kar2022interpretability}. These include but are not limited to, cortico-cortical recurrence, thalamocortical loops, basal ganglia loops, cortical laminar structure and local circuitry, cell types, biophysically grounded dendritic and neuronal models, synaptic dynamics and adaptation, spiking mechanisms, etc.  It is currently unclear how much these structures will turn out to be critical for accurately closing the accuracy gaps in predicting the behaviors supported by the ventral stream or predicting functional neural measurements along the ventral stream (Section 2.1). However, one fundamental principle in neuroscience is that form, encompassing morphology and anatomy, invariably constrains function. By this logic, enhancing SMART models to more closely mirror these recognized anatomical facts will likely bolster their empirical functional accuracy.

\begin{textbox}[h]
    
\section{Box 5: Role of AI engineering in systems neuroscience?}

%The ImageNet challenge \citep{russakovsky2015imagenet} demonstrates the transformative role large-scale datasets and benchmarks play in computer vision's progress as a field. Even today, the ImageNet challenge remains a standard against which the efficacy of computer vision models is gauged. Drawing a parallel in the realm of visual neuroscience, there is a growing recognition of the critical role publicly accessible large-scale datasets play, as well as the imperative for reproducible evaluation and adjudication of alternative computational models of vision poised to predict and explain such data.

%As discussed in this article, there has been a healthy collaboration of ideas between developing models that can perform visual problem-solving for computers and the discovery of brain mechanisms that enable primate vision. %With the recent emergence of highly accurate ANN models of object recognition, in the recent times, most primate vision models have been inherited from computer vision. Interestingly, 
It is a striking observation that AI engineering to performance optimize a ventral-stream inspired family of deep ANN models -- \emph{but without further regard for the brain} -- returned our currently leading neuroscientific models of the brain mechanisms \citep{yamins2014performance, cadieu2014deep}.   Does this mean that neuroscientists just sit back and wait for AI engineering to deliver even better neuroscientific models?  In theory, it should be obvious that this trend must have limits -- one cannot model all of the biology without empirically studying the biology.  Indeed, while this remarkable upward trend continued for SMART models of the ventral stream from 2013-2016, we have already seen the turning point (Fig. 7).  Today, more accurate neuroscientific SMART models are deriving from a close \emph{collaboration} of natural sciences experiments \emph{and} AI engineering. However, other areas of systems neuroscience are, we believe, still on the upward trajectory in that even loosely brain-inspired AI engineering is producing the leading neuroscientific models \citep{schrimpf2021neural,kell2018task}.  Given the history of visual neuroscience, we expect those trends to continue and then evolve in a similar way.

%It’s necessary to acknowledge that this trajectory, while extremely promising, isn't inexhaustible. We should not expect this trend to continue indefinitely. In fact, recent trends in Brain-Score (\textbf{Figure 7}) already indicate the emergence of this trend. Given that the primary goal of computer vision systems is to achieve the most robust and accurate behavior, far surpassing the human visual system, such a trend is not unlikely to occur. Primate visual neuroscience aims to discover algorithms that help simulate a visual system similar to the primate brain that could demonstrate similar internal mechanisms and behavioral error patterns. Thus we call for a closed loop interaction between modelers and experimentalists to inherit the advances from AI but reshape it to improve models of the brain. 

\end{textbox}

\begin{figure}[h]
\includegraphics[width=13 cm]{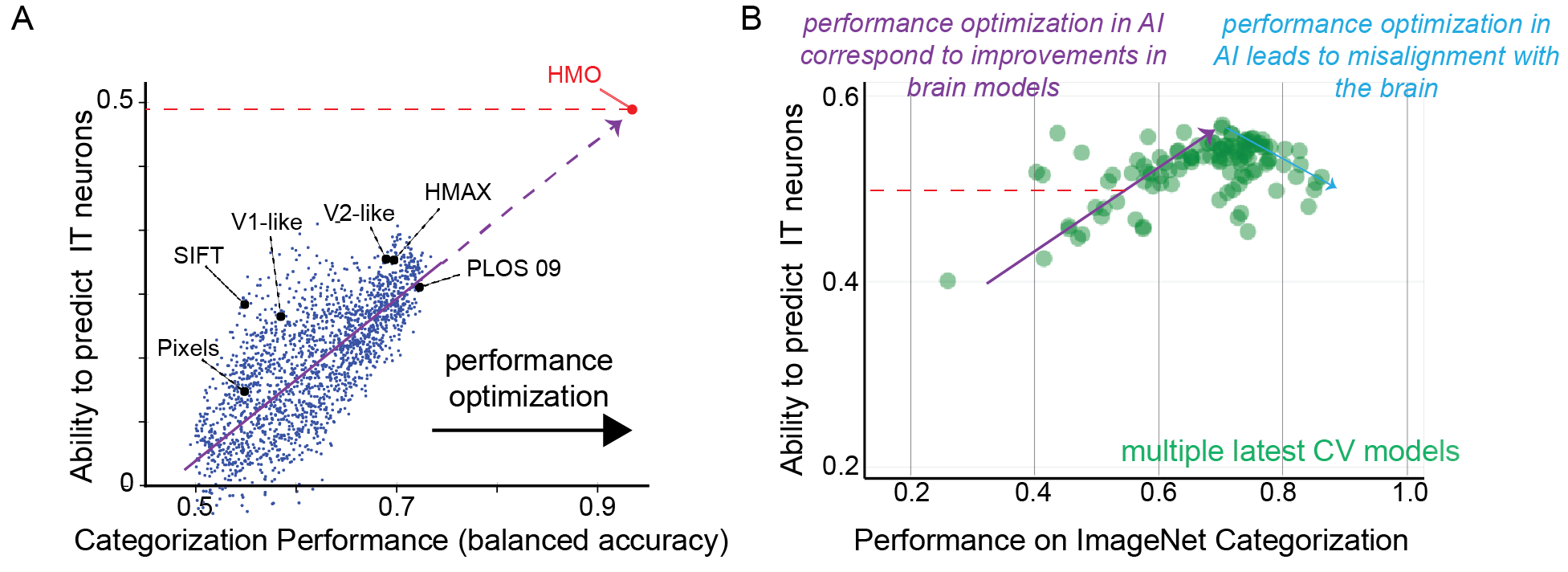}
\caption{ \textbf{Relationship between object categorization performance and neural alignment of ANN models} \textbf{A.} Each dot is a set of CNN model features.  X-axis: performance on an object categorization task (not ImageNet).  y-axis:  IT neural predictivity of the model features (adapted from  \citep{yamins2014performance}). Noting this trend (A), Yamins et al. \citeyear{yamins2014performance} used optimization methods to develop the HMO model, and found that HMO's penultimate layer (red dot) explained unprecedented levels ($\sim$ 50 $\%$) of IT response variance at that time. \textbf{B.} Relationship between  Object categorization accuracy (ImageNet \citep{russakovsky2015imagenet} accuracy) and IT predictivity (see Box 5, results taken from Brain-Score). Beyond HMO (A), improvements in the categorization performance of the overall model continued to produce a hidden feature layer that followed the trend that Yamins et al. \citeyear{yamins2014performance} had identified, leading to even higher IT predictivity (purple line).  However, the trend did not empirically continue past 2017 (see Box 5 for perspective on this).}
\label{fig7}
\end{figure}

The challenge of computationally integrating all these elements remains formidable. As a result, researchers are taking a piecemeal approach, examining the impact of each omitted or inaccurately represented element individually. A relevant query in this context is: How does the incorporation of cortico-cortical recurrence augment the functional accuracy of SMART models?  Many studies \citep{kar2019evidence,kietzmann2019recurrence,tang2018recurrent, del2016adaptation, joukes2014motion} have motivated the potential shortcomings of a feedforward-only approach to modeling primate vision. These studies highlight the need to incorporate recurrent computations to produce a more accurate alignment of ANNs with the primate visuocortical processing machinery. This has led to the development of deep recurrent neural networks \citep{kubilius2019brain, nayebi2021goal, zamir2017feedback} that are still mostly “work in progress” in terms of their overall performance and predictive capability of neural responses. Some of these networks (e.g., CORnet-S; Kubilius et al. 2019) indeed predict neural benchmarks much better than their feedforward predecessors. Another recent study \citep{cornford2020learning} underscores the disparities between ANNs and biological networks, highlighting the absence of Dale’s principle \citep{dale1935pharmacology} in ANNs — a principle ensuring that biological neurons are either exclusively excitatory or inhibitory. The study introduces Dale's ANNs (DANNs), which, inspired by feedforward inhibitory interneurons, incorporate separate excitatory and inhibitory units without compromising learning performance. Other studies \citep{blauch2022connectivity, } have also emphasized the importance of incorporating realistic connectivity constraints within the ANNs to better align them with biological functions. 

Given the depth and intricacy of the augmentations possible in the SMART models, it becomes imperative to underscore the critical importance of experimental evaluation. To truly gauge the accuracy of these models, future experiments must be equipped with cutting-edge tools and methodologies that not only test their predictions but produce data at a grain that can be utilized to refine and improve the models. As we move forward, the nexus between the predictions of the next-generation advanced SMART models and empirical evidence will be the cornerstone of our understanding. However, it is important to remind the reader that the endeavor of integrating new neuroscientific components into SMART models can serve beyond overall improvements in brain alignment against the measures the field already has. There's another dimension of value to consider. Even if no quantitative empirical accuracy gains are realized, the incorporation of these components provides routes that could allow for novel perturbation and control tests (see 2.2) that might reveal new clinical translation opportunities (further discussed below).

    \section{Possible applications of SMART models}

In this review, we underscore how a rational quest for SMART models of object recognition has led the field to particular, brain-mapped ANNs as the currently leading candidates (Section 2).    We also pointed out that these leading SMART models are not 100\% empirically accurate in the functional neural and behavioral comparisons that have been formally examined thus far (Section 3), and they are likely not 100\% accurate in other similar comparisons that have not yet been formally made, and they are not yet at the level of subcellular and molecular mechanisms (type iv, Section 1).   

Looking beyond these scientific challenges that will surely be solved as SMART models continue to evolve and improve, in this final section provide perspective on a philosophical challenge that SMART models have engendered.  Specifically, these large-scale, integrated models (which are all deep ANNs) have been referred to as uninterpretable ``black boxes" in both neuroscience and in AI/computer vision. Unlike the brain, these models are fully observable, so the black box criticism is, in this sense, inappropriate. 
%Nevertheless, while these models, like humans and monkeys,  achieve impressive behavioral performance in object recognition and other related visual tasks and impressive alignment with the internal processing of the ventral visual stream, 
Nevertheless, we agree that it can feel impossible to look inside a current leading SMART model of object recognition and intuitively understand or interpret how it arrives at its behavioral decisions, to understand the image conditions in which it will succeed or fail, and to understand the optimization conditions that will produce similar arrangements of mechanisms.

In neuroscience, this criticism has resonance because if the goal is to use SMART models as a proxy for our understanding of the brain's visual processing, then weaknesses in interpretability seem like limitations. More succinctly, if we succeed in building a digital twin (i.e. a SMART model) but do not fully understand it in all the ways outlined above, how can we say that we understand the brain system that it purports to explain?   We expect that theoretical approaches that examine fully observable SMART models, rather than the brain itself, will help our field close some of these gaps \citep{cohen2020separability, poggio2020theoretical}. But what if that does not -- or cannot -- happen to our satisfaction?

In this review, we first engaged this criticism by logically articulating the concept of "mechanistic understanding" (Section 1), and we note that many celebrated mechanistic models in neuroscience are subject to similar criticisms (e.g. the Hodgin-Huxley model of action potentials, which is non-linear and not always intuitively predictable).  But here we forward another even more important answer to this criticism: beyond our field's quest for scientific understanding is a pragmatic goal -- to improve the quality of human life.  And, as we outline next (paraphrased from \citealp{schrimpf2020integrative}), even not-easy-to-interpret SMART models will almost surely be capable of guiding us to new ways to do that. 

\subsection{Research applications}
In the ventral visual stream, SMART models are already being used to focus experimental resources on the most interesting aspects of brain function that are not yet accurately described. For example, by drawing on the predictive accuracy of these models, neuroscientists can now use them to control individual neurons and entire populations of neurons deep in the visual system via model-synthesized patterns of light applied to retinae \citep{bashivan2019neural, ponce2019evolving}.  
Such model-driven stimulus synthesis methods can be used to better adjudicate among alternative SMART models  \citep{golan2020controversial}. Similarly, variants of SMART models that predict image memorability can be used to discover image manipulations that causally affect human memory performance \citep{goetschalckx2019ganalyze}.

In another study, SMART model were used to discover that the macaque IT cortical responses are surprisingly sensitively to small, \emph{model-guided} image perturbations \citep{guo2022adversarially}, and that human category judgements are also surprisingly sensitively to small \emph{model-guided} image perturbations  \citep{gaziv2023robustified}. Given the vastness of image space, this previously unknown neurobiology and previously unknown perceptual sensitivities would have been impossible to discover without those models. Indeed, these discoveries trace back to theoretical and empirical analyses of the SMART models themselves \citep{goodfellow2014explaining} and efforts to build new candidate SMART models \citep{madry2018towards}

Stepping back, one can see the overall research trend here is this: Step 1) Neuroscientists, cognitive scientists, and computer scientists work together to transfer the structure of a brain subsystem (which is only partially measurable) into one or more SMART models, where reasonable choices of task and optimization methods help fill in much of the non-measurable model structure. Step 2) They then use those now fully observable "digital twin" SMART models to make and test predictions about that brain subsystem, leading to new scientific discoveries and exposing new model-vs-brain mismatches.  Step 3) Use those models (not the brain itself!) to build a deeper theoretical understanding, which then leads to new SMART models (i.e. a new run of Step 1).  Repeat the cycle.  

\subsection{AI applications} As our field discovers SMART models that are ever more closely aligned to primate brains and human behavior, we will in fact be discovering machine systems that, for example, successfully generalize more like humans, are less susceptible to adversarial attacks, and potentially more energetically efficient. For more general reviews on this topic, we refer the reader to \citealp{hassabis2017neuroscience}.

\subsection{Brain-machine interface applications} Sufficiently accurate SMART models of visual processing can be used to determine complex, non-intuitive direct brain stimulation patterns \citep{chen2020shape,azadi2023image} that could be applied in mid- and high-level visual areas to replicate visual percepts (e.g. in blind individuals). 

\subsection{Brain disorders applications} For most brain disorders, the treatment goal is to precisely modulate brain activity in a beneficial way. While it is commonly assumed that such interventions will be delivered via new pharmaceuticals (difficult to target precisely) or perhaps with inserted probes (dangerous and still not precise), accurate SMART models might reveal entirely new treatment possibilities. For instance, by directing the synthesis of patterns of light delivered to the retina that predictably and precisely modulate entire populations of neurons deep in the brain at single-neuron resolution to, in turn, beneficially improve cognitive states such as anxiety or depression. 

To pursue this and other such interventions, a concerted effort is needed to develop and fine-tune models that cater to individuals with unique neurological and behavioral challenges. These refined models, when validated, could revolutionize clinical interventions. A reasonable speculation is that models that are better at predicting the neurotypical brain and behavior will be more useful in the clinic. Model generators that can explain human \emph{variation} will be needed to unlock that utility. 

In addition, the more control knobs (aka neural mechanisms) that a model can engage with, the more potential clinical interventions it can provide. This ties back to section 1.3 underscoring the imperative to build SMART models that connect to the cellular and molecular components (type iv SMART models) to unlock powerful molecular and genetic interventional toolboxes. 

%Indeed, the presence of sophisticated tools that can interface with certain neural mechanisms could shift the priority toward their inclusion in SMART models, potentially sidestepping the behavior-based hierarchy of model motif integration we previously proposed.

\subsection{The future}

The above are just a few application areas in the context of SMART models of the ventral stream and visual object recognition.  We hope the reader will see how these same application themes can -- and we think will -- readily generalize to SMART models of other aspects of cognition such as audition \citep{kell2018task}, language \citep{schrimpf2021neural}, motor planning \citep{rajalingham2022recurrent}, and beyond.  The key overall point is that all of the above applications – and myriad others not yet imagined – will get ever better with ever more accurate SMART models, \emph{even if we do not fully understand those models in all the ways that we aspire to}. 

With ample incentive and progress in building ever-more accurate SMART models of visual perception and other areas of cognition, we are now on the cusp of realizing some of the application gains above. This is both an enormous opportunity and a challenge.  Ethical quandaries arise with the scientific capability to influence behavior—how do we harness such capability responsibly, ensuring its use predominantly benefits individuals without potential harm? Such research programs must not only navigate the technical challenges but also spearhead discussions on the ethical deployment of such potent models in all application arenas.

% Summary Points
\begin{summary}[SUMMARY POINTS]
\begin{enumerate}
\item \textbf{Breakthrough: }The past decade has seen ventral-stream-inspired deep ANNs emerge that achieved unprecedented, human-like accuracies on core object recognition tasks.

\item \textbf{Prediction: } Specific ANNs, once mapped to the brain (SMART models), have internal neural representations that surprisingly mimic activity along the primate ventral visual pathway; And similarly mimic primate core object recognition behavioral patterns.

\item \textbf{Control: } current leading SMART models can be used to synthesize goal-directed stimuli that successfully modulate targeted neural populations in non-trivial ways.

\item \textbf{Glass half empty:} While current ANN models provide an advanced understanding, they do not yet fully capture the neural and behavioral nuances of the ventral visual processing stream and its supported core object recognition behaviors.

\end{enumerate}
\end{summary}

% Future Issues
\begin{issues}[FUTURE ISSUES]
\begin{enumerate}

\item \textbf{Modelers --Call to action: } Create next-generation SMART models of the ventral visual stream, ensuring tighter integration with neural and behavioral experiments and corresponding application gains.
\item \textbf{Expand behavioral domains: }Explore and evaluate alternative SMART models that expand beyond visual object recognition to encompass broader aspects of visual intelligence.
\item \textbf{Bridge to even finer spatial scale:} Delve into cellular-level implementations of SMART models, integrating known anatomical details like cortico-cortical recurrence, cell types, and synaptic dynamics.
\item \textbf{Responsible AI:} Ensure that ethical considerations are at the forefront as the field begins to leverage this new SMART-model-based understanding for potential clinical interventions and behavioral modulation.
\end{enumerate}
\end{issues}

%Disclosure
\section*{DISCLOSURE STATEMENT}
The authors are not aware of any affiliations, memberships, funding, or financial holdings that might be perceived as affecting the objectivity of this review. 

% Acknowledgements
\section*{ACKNOWLEDGMENTS}

KK was supported by the Canada Research Chair Program, Simons Foundation Autism Research Initiative (SFARI), Google Research Award, and the Canada First Research Excellence Funds (VISTA Program). JJD was partially funded by the Office of Naval Research (N00014-20-1-2589, JJD), MURI N00014-21-1-2801 (JJD), the National Science Foundation (2124136, JJD), the Simons Foundation (542965, JJD),  NC-GB-CULM-00002986-04, and the Semiconductor Research Corporation (SRC) and DARPA.
We thank Martin Schrimpf, Carlos Ponce, and Will Xiao for sharing editable drafts of Figures, and helpful discussions.

% References
%
% Margin notes within bibliography

\bibliography{refs.bib}

\end{document}